\documentclass[a4paper,notitlepage,10pt]{article}
\usepackage{amssymb,amsfonts,amstex,axodraw}
\usepackage{latexsym}
\usepackage{newcent}
\newtheorem{Thm}{Theorem}[section]

\newtheorem{Def}{Definition}[section]

\newcommand{\R}{\mathbb{R}}
\newcommand{\V}{{\bf V}}

\newcommand{\N}{\mathbb{N}}

\begin{document}

\title{Finite Renormalizations in the Epstein Glaser Framework
and Renormalization of the $S$-Matrix of $\Phi ^4$-Theory}

\author{G. Pinter}

\date{}

\maketitle

\vspace{-1.1cm}
\begin{center}
\textsc{%
II. Institut f\"ur Theoretische Physik\\
Universit\"at Hamburg\\
Luruper Chaussee 149\\
22761 Hamburg 
Germany}\\
e-mail: \verb+gudrun.pinter@desy.de +
\end{center}

\vspace{1cm}
\begin{abstract}
\noindent
A formula describing finite renormalizations is derived in
the Epstein-Glaser formalism and an explicit calculation of finite 
counterterms in $\Phi ^4$-theory is performed.
The Zimmermann identities and the
action principle for changes of parameters in the interaction
are presented independent of the adiabatic limit.
\end{abstract}


\section{Introduction} 

Renormalization is an old art of removing divergences which occur
unavoidably in QFT.
In course of time many calculational techniques, more or less mathematical,
were developed.
Usually the more mathematical formulations of renormalization
were too abstract for practical purposes, like the Epstein Glaser
approach \cite{epstein} of renormalization. Apart from the work
of Scharf \cite{scharf} and Stora \cite{stora} \cite{sto} 
nothing was done in
this framework for a long time. Nevertheless a further development
of the Epstein Glaser method is worth wile, because it turned out
that this method is best suited for the construction of theories
on curved space times \cite{fredneu} \cite{mdf}. 
Its advantages are the local character 
and the formulation in position space.

Another more abstract formulation of renormalization theory is the BPHZ-
renormalization. In this framework some fundamental results of the 
structure of renormalization were achieved, namely the forest formula
and the action principle. The latter describes how Green's functions
change by a variation of parameters in a theory. In the derivation of
the action principle Lowenstein
\cite{low} used the Gell-Mann-Low formula to express
Green's functions in terms of free fields. 
In this context the action principle
is a consequence of simple properties of free field insertions 
into the $S$-matrix. In the present work we call these properties action
principle, because they describe the underlying basic structure and
can be proved independently of the adiabatic limit.
This is justified by the fact that the Gell-Mann-Low formula is valid
in Epstein-Glaser renormalization in the adiabatic limit, if this limit
can be performed. Therefore in the adiabatic limit our action principle
indeed describes via the Gell-Mann-Low formula the dependence of Green's 
functions on parameters of the theory.
D\"utsch and Fredenhagen \cite{neu} give another derivation of the
action principle. In contrast to our derivation they use insertions
into time ordered products
($T$-products) of interacting fields. Using the fact that interacting
fields are up to a factor insertions in the $S$-matrix one can transform
the two formulations into each other.
In their comparison with
the usual action principle they only consider variations of
the interacting part of the Lagrangian with mass dimension 4. In this
case their action principles coincides with the usual one 
in the adiabatic limit.
Breitenlohner and Maison \cite{maison} 
succeeded in formulating the action principle also in dimensional 
renormalization.

In this work we will give a formulation of finite renormalizations
in the Epstein Glaser approach corresponding to the forest formula.
Furthermore we give a formulation of $T$-products and insertions
in the Epstein Glaser formalism so that the derivation of the part
of the action principle concerning changes in the interaction
is analogous to that of Lowenstein \cite{low}.
To fill the
gap between the theoretical formulation and practical calculations
we demonstrate how to renormalize the $S$-matrix in $\Phi^4$-theory
up to the third order.

Many calculations concerning renormalization can be found in the
book of Zavialov \cite{zav}. Often they are similar to the results 
presented here, but they are not formulated in the sense of distributions.
Especially the structure of the renormalization presented in \cite{zav}
corresponds to the formulas given in the theorems (6.1) and (6.2).

To make this work selfcontained, we sometimes repeat some
results of other articles in a form fitting to our calculations.
We hope it is helpful for the reader to get a better understanding of
the method.

This work is divided into five sections. In the first one
we briefly repeat the basics of causal perturbation theory 
in the framework of the Wightman axioms. 

In the next section we present
a mathematical description of the time ordered
product. The inductive construction of Epstein and Glaser \cite{epstein}
is described.
The main result of this section is the description of finite 
renormalizations by a family of functions $\Delta _n$. 

Section 4 prepares the calculations of section 5
and contains a short summary of \cite{scharf} \cite{epstein} \cite{fred}
\cite{fredneu}.
After some microlocal analysis we see that renormalization is 
nothing else than an extension of distributions on an 
appropriate space of test functions. We repeat the Lorentz invariant
form of the extension presented in \cite{wir} 
which is used in the calculations
and discuss some properties of the $W$-operator.

In section 5 we show how renormalization of
the $S$-matrix of $\Phi^4$-theory works up to
the third order. In the results we only list the terms surviving
in the adiabatic limit, but the calculation
can be performed without using the existence of this limit.
In contrast to other renormalizations in momentum space
the complexity grows not with the number of loops but of vertices
in a diagram. Thus the second order calculation is simple because
subdivergences first appear in the
third order calculation. To come back to the abstract
formulation of the Epstein Glaser approach we list the normalization
conditions for a scalar theory. Some of them are an abstract form of
the rules used in the calculations. With the Gell-Mann-Low formula
we derive the Dyson Schwinger
equations (DSE) from the normalization condition N4 in $\Phi ^4$-theory.

In the next section we derive the action principle for variations
of parameters in the interaction analogously to \cite{low}. 
The first subsection is based on the theorem of
perturbative renormalization theory \cite{stora}. Using the form
of finite renormalizations of section 2 we are able to determine
the counterterms in the Lagrangian compensating
a change of renormalization explicitly. Then we define insertions
into $T$-products and show that they have
some of the properties of the insertions of \cite{low}. 
Relations between insertions of different 
degrees are described by the Zimmermann identities \cite{zim}, that are
formulated in the framework of Epstein Glaser renormalization.
Finally we prove the formulation of the part of
the action principle concerning changes in the interaction in terms
of insertions in the $S$-matrix independent of the adiabatic limit. 
The action principle for changes in the parameters of the free 
Lagrangian turns out to be more complicated, so we postpone it.

These results and the missing part of the action principle
allow a derivation of the renormalization group
equations (RGE) as in \cite{low}.
The RGE and the DSE are also valid outside perturbation theory.
They provide for an important tool for nonperturbative constructions
and methods, see for example \cite{zj} and \cite{stingl}.
The action principle is used in algebraic renormalization \cite{psor},
which leads to a systematic renormalization of the standard model
of electroweak interaction to all orders of perturbation theory
\cite{ekraus}.
This work is a first step of a translation of these
methods into the Epstein Glaser formulation of renormalization.

We always use the manifold $\R^4$ with the Minkowski metric for 
spacetime. With the work of \cite{fred}, \cite{123},
\cite{fredneu}, \cite{marek} it should be easy to formulate the
arguments on curved space times, often one only has to replace
$\R^4$ by an arbitrary Lorentz manifold $\mathcal{M}$.

\section{Causal Perturbation Theory and Epstein Glaser Renormalization}

A mathematical precise formulation of a QFT was given 
by G\r{a}rding and Wightman \cite{gw}. They treat a QFT as a tupel

\begin{eqnarray}
\left( \mathcal{H}, U, \phi, D,  |0 \rangle \right)
\end{eqnarray}

of a separable Hilbert space $\mathcal{H}$, a unitary representation
of the restricted Poincar\'{e} group $\mathcal{P}_+^{\uparrow}$,
field operators $\phi$, a dense subspace $D$ of  $\mathcal{H}$ 
and the vacuum vector 
$ |0\rangle$. The tupel has to fulfill the Wightman axioms \cite{sw}. 
Among others they state
that the fields $\phi$ are local operatorvalued distributions, 
welldefined on the dense domain $D$ of the Hilbert space. $D$
contains the vacuum. In the construction of the $S$-matrix
it is sufficient to regard the dense domain $D_0 \subset D \subset
\mathcal{H}$, generated by all vectors which can be constructed by
applying a finite set of field operators to the vacuum:

\begin{eqnarray}
D_0 = \{ \psi \in \mathcal{H} | \psi \in \langle \phi (f_1) \ldots
\phi (f_n) |0 \rangle , n \in \N \rangle \}.
\end{eqnarray}

Therefore any formulation of perturbation theory has to take 
into account the distributional character of the fields and interaction terms,
and it has to make sure that the operators are well defined on $D$ 
(or in the case of the $S$-matrix on $D_0$). Causal perturbation
theory is in the sense of these axioms,
because it has all those properties.
It was founded by the work of
Stueckelberg \cite{stue}, Bogoliubov and Shirkov \cite{meister}.
In their formulation every interaction term is accompanied by a test function 

\begin{eqnarray}
g \in \mathcal{S} \left( \R ^4 \right),
\qquad \qquad \qquad \qquad 
g  : \R ^4 \longrightarrow \left[ 0,1 \right]
\end{eqnarray}

which switches the interaction on and off at different space time points
($\mathcal{S}$ is the Schwartz space of testfunctions).
$g$ vanishes at infinity and therefore provides for a 
cutoff for long range interactions.
All quantities of the causal construction are formal power series in this
function, especially the $S$-matrix is constructed with the following
ansatz:  

\begin{eqnarray}
 S \left( g \right) = 1 + \sum_{n=1} ^{\infty} \frac{i^n}{n!}
\int d^4 x_1 \ldots \int d^4 x_n \ S_n \left( x_1, \ldots, x_n \right)
  g \left( x_1 \right) \ldots  g \left( x_n \right). \label{S}
\end{eqnarray}

The coefficients
$S_n$ are operatorvalued distributions, smeared with the
switching function $g(x)$, such that each term in the sum is a
welldefined operator on $D_0$.

Epstein and Glaser \cite{epstein} have found that the $S_n$
can be determined by a few physical properties of the $S$-matrix, namely

\begin{enumerate}

\item Translation covariance

\item Lorentz covariance

\item Unitarity

\item Causality

\end{enumerate}

and the renormalization conditions.
Actually for the inductive construction of the $S_n$ proposed by
Epstein and Glaser only causality is important. To prove
that the resulting
$S(g)$ is a welldefined operator on $D_0$ they apply Wick's
theorem and need translation covariance. In \cite{fredneu}
it is shown how translation covariance can be substituted by a
condition on the wave front sets of the $S_n$. This makes the method
work on curved space times, too. The other properties are 
realized by the normalization conditions.

With some calculations, the properties above are transfered to the
following properties of the coefficients $S_n$:

\begin{enumerate}

\item Translation covariance: 
$S_n \left( x+a \right) = U \left( a \right)
S_n \left( x \right) U^{-1}  \left( a \right)$.

\item Lorentz covariance:
$S_n \left(  \Lambda  x \right) = U \left( 0,\Lambda
\right) S_n (x)  U^{-1}  \left( 0,\Lambda  \right)$.

\item In the following, we will often work with sets of indices.
We will denote with $J=\{1, \ldots,n\}$ the full set of indices
and with $I$ any subset of $J$. For  $I= \{ i_1,\ldots, i_k \}$
we use the notation

\begin{eqnarray}
S_k \left(  x_j| j \in I \right)
:= S_k \left( x_{i_1}, \ldots,  x_{i_k} \right).
\end{eqnarray}

From the unitarity condition the following equation is obtained:

\begin{eqnarray}
\sum _{ I\subset J \atop |I| = k}
S_k \left(  x_i| i \in I \right) S^{+}_{n- k}
\left(  x_j | j \in J \setminus I \right) =0.
\end{eqnarray}

\item Causality of the $S$-matrix yields the factorization property:

\begin{eqnarray}
S_n \left( x_1, \ldots, x_n \right) =  S_i
 \left( x_1, \ldots, x_i \right) S_{n-i} 
\left( x_{i+1}, \ldots, x_n \right) 
\label{fact}
\end{eqnarray}

for $\{ x_1, \ldots, x_i \}
\gtrsim \{ x_{i+1}, \ldots, x_n \}$, where $x \gtrsim y$ means that
$y$ does not lie in the forward light cone of $x$.

This property implies the locality of the $S_n$:

$\left[S_n \left( x_1, \ldots, x_n \right),
 S_m \left( y_1, \ldots, y_m \right) \right]=0$
if all the $x_i$ are spacelike to the $y_j$.

\end{enumerate}

The factorization property (\ref{fact})
is used to identify the coefficients
$S_n$ with timeordered products. If we further demand
$S_1= \mathcal{L}_{int}$ the higher $S_n$ can be constructed inductively
\cite{epstein}.
This is described in the next section.

From the $S$-matrix the interacting fields can be obtained by
the following formula of Bogoliubov \cite{meister}:

\begin{eqnarray}
\phi_g (h) = \left. 
\frac{d}{d \lambda} S^{-1}(g) S(g+ \lambda h)\right|_{\lambda =0}.
\end{eqnarray}

They are formal power series in the switching function $g$ like the
$S$-matrix. To get rid of this function at the end of the construction
the adiabatic limit $g(x) \to 1$ has to be performed. This limit
bears some problems, because all infrared divergences appear which were
avoided in the local formulation.
By definition the adiabatic limit exists in the weak sense if
all Green's functions exist in the sense of tempered
distributions for $g \to 1$ choosing suitable normalizations. 
If it is further shown that for $g \to 1$ the $S$-matrix
is a unitary operator one says that
the adiabatic limit exists in the strong sense.
The adiabatic limit is constructed by choosing a sequence of
functions $g_n$ with $\lim_{n \to \infty}g_n =1$, one has to be
careful because sometimes the
result depends on the choice of the $g_n$.

The weak adiabatic limit was proved to exist for massive theories 
\cite{epstein},
QED and massless $\lambda :\Phi ^{2n}:$ theories \cite{abc}. The existence
of the adiabatic limit in the strong sense has only been proved for 
massive theories \cite{ep2}.

\section{The Time Ordered Product} \label{tprod}

For a precise mathematical formulation, the $T$-products introduced
in this section are not defined as usual on Wick monomials of quantized 
fields. This idea was introduced in \cite{fm}.
Let $\mathcal{A}$ be a commutative algebra generated by the
so called classical symbolical 
fields $\phi_i$ and their derivatives. They are called
symbols, because there is no relation like the Klein-Gordon equation,
so that the fields and their derivatives are
linearly independent.
We regard

\begin{eqnarray}
\mathcal{D} \left( \R^4, \mathcal{A} \right) \ni f =
\sum_i g_i (x) \phi _i 
\end{eqnarray}

where the sum is a finite sum over elements $\phi_i$ of the
algebra $\mathcal{A}$.
An element $f$ is given by its coefficients $g_i \in \mathcal{C}
^{\infty}_0 (\R^4)$.

The time ordered product ($T$-product) is a family of maps
$T_n, n \in \N$, called $T_n$-products. They are functions from
$\left( \mathcal{D} \left( \R^4, \mathcal{A} \right) \right)^{\otimes n}$
into the operators on $\mathcal{H}$ with the following properties:

\begin{enumerate}

\item $T_0 = 1$
 
$T_1 \left( f \right)= \sum_i :\phi _i (g_i): \qquad \forall \quad
f \in \mathcal{D} \left( \R^4, \mathcal{A} \right),$

where the sum is over all generators of $\mathcal{A}$.
Each local field is the image of an element $f \in 
\mathcal{D} \left( \R^4, \mathcal{A} \right)$ under $T_1$.
We define the $T$-products such that $T_n \left( g (d\phi)\right)
= T_n \left( (d ^t g)\phi \right)$ is fulfilled for derivatives $d$
up to the second order. 
$T_1$ is not injective, because the Wick products obey the wave equation.
For example, in a free scalar theory we obtain 

$T_1 \left( g \Box \phi 
+ g m^2 \phi \right) = : \phi \left( \left(\Box + m^2 \right) g
\right): =0$.

\item Symmetry in the arguments:  

\begin{eqnarray}
T_n \left(  f_1, \ldots ,f_n \right) =
T_n \left(  f_{\pi_1},  \ldots , f_{\pi_n} \right) 
 \qquad \forall \  \pi \in S_n \qquad 
\forall \ f_i \in \mathcal{D} \left( \R^4, \mathcal{A} \right),
i=1, \ldots n,
\end{eqnarray}

where $S_n$ is the set of all permutations of $n$ elements.

\item The factorization property:

\begin{eqnarray}
T_n \left(  f_1,  \ldots ,f_n  \right) =
T_i \left(  f_1, \ldots ,f_i \right)
T_{n-i} \left(  f_{i+1}, \ldots ,f_n \right) \label{fac}
\end{eqnarray}

if \
$\mathrm{supp} f_1 \cup \ldots \cup \mathrm{supp} f_i \gtrsim 
\mathrm{supp} f_{i+1} \cup \ldots \cup \mathrm{supp} f_n$
and $f_i \in \mathcal{D} \left( \R^4, \mathcal{A} \right) \ \forall \ i$.

\end{enumerate}

We remark that the $T_n$ are multilinear 
functionals in the arguments $f_i$, for instance
$T_2( f_1 + f_2 , f_3 )
= T_2( f_1, f_3)+ T_2( f_2, f_3).$
Therefore it suffices to know their values
for one kind of interaction term $f$ (polarization identity),
and we can omit the indices of the $f$.
 
We now want to
construct higher $T_n$-products with the help of the factorization
identity as expressions of lower $T_n$-products. 
This is possible if $\cap_{i \in J} \mathrm{supp} f_i = \emptyset$ with
$J = \{1, \ldots,n\}$, in other words, the total diagonal

\begin{eqnarray}
D_n = \{ (x_1, \ldots, x_n) \in \R^{4n} | x_1 = x_2 = \ldots x_n \}
\end{eqnarray} 

is not in the support of the tensorproduct of the $f_i$.
In the calculations this is achieved
by multiplication of the distribution with a causal partition of
unity:

\begin{Def}
A causal partition of unity in $\R ^{4n} \setminus D_n$ is a set of
$\mathcal{C} ^{\infty}$-functions 

$p_I^{(n)}:
\  \R^{4n} \setminus D_n \longrightarrow \R$ with the following properties:

\begin{enumerate}

\item $\mathrm{supp} p_I^{(n)} \subset 
\{ (x_1, \ldots, x_n) \in  \R^{4n} \setminus D_n \ 
| \ x_i \gtrsim x_j \ \forall i \in I, j \in I^c\}$

\item  
$\sum_{ I  \subsetneq J \atop I \not= \emptyset}
\left. p_I^{(n)} \right| _{\R^{4n} \setminus D_n} = 1.$

\end{enumerate}

\end{Def}

Since $\mathrm{supp} f_1 \otimes \ldots \otimes f_n$ is contained in a
compact region, $\mathrm{supp} f_1 \otimes \ldots \otimes f_n \cap
C^I$ is contained in a compact region, where 
$C^I = \{ (x_1, \ldots , x_n)\subset \R^{4n} | 
\ x_i \gtrsim x_j \ \forall i \in I, j \in I^c\}$.
In this region the part $p_I^{(n)}$ of the partition of unity
can be written as a finite sum of factorized terms
$p_I^{(n)}= \sum_k p_{I,k}^1 (x_1) \cdot \ldots \cdot p_{I,k}^n (x_n)$.
Then a $T$-product is constructed according to the factorization
property as follows:

\begin{eqnarray}
T_n^0 \left( f_1, \ldots ,f_n \right) =
T_n^0 \left( \bigotimes _{k=1}^n  f_k \right) 
= \sum_{I \subsetneq J \atop I \not= \varnothing} \sum_k
T _{|I|}\left( \bigotimes _{j \in I} 
p_{I,k}^j f_j^I \right)
T _{|I^c|} \left( \bigotimes _{l \in I^c}  p_{I,k}^l f_l^I \right).\label{ind}
\end{eqnarray}

The last step in this construction of Epstein and Glaser
\cite{epstein} is the extension of the
right hand side of (\ref{ind}), if the support of the
tensor product of the $f_i$ contains the total diagonal $D_n$.
We call this shortly the extension of the $T$-product to the
total diagonal.
This extension is not unique, so there are several 
$T$-products differing from another by finite renormalizations.
In renormalization theory this corresponds to the free choice of the
renormalization constants. 

Now we can describe the structure of finite renormalizations with the
following theorem, which arose from discussions with K. Fredenhagen:
(it can be seen as the precise formulation of a formula given by
\cite{meister})

\begin{Thm} \label{meinsatz}

Let $T, \hat{T}$ be two different $T$-products.
Then there are functions $\Delta_n : \mathcal{D} \left( \R^{4n}
,\mathcal{A}^n \right) \to \mathcal{D} \left( \R^{4},\mathcal{A} \right)$
with $\mathrm{supp} \ \Delta_n \subset D_n$ and

\begin{eqnarray}
\hat{T}_n \left( \mbox{$\bigotimes$} _{j \in J} f_j \right) =
\sum_{ P \in Part(J) }
T_{|P|} \Bigl[ \bigotimes_{O_i \in P}
\Delta_{|O_i|} \left( \mbox{$\bigotimes$}_{j \in O_i} f_j \right)
\Bigr]. \label{msatz}
\end{eqnarray}

\end{Thm}

For the interpretation of the operatorvalued distribution $\Delta_n$
we go back to equation (\ref{ind}). 
The lower $T$-products on the right hand side of (\ref{ind}) 
consist themselves
of products of lower $T$-products which were extended to a subdiagonal
of $\R^{4n}$. We call the extension to these subdiagonals the 
{\it renormalization
of subdivergences} and the extension to the total diagonal in the last
step the {\it renormalization of the superficial divergence}.
 
Fixing the extensions of all $T_n$-products to the diagonals
defines us
a special $T$-product. The distribution $\tilde{\Delta}_n
= T^{-1}_1 (\Delta_n)$ is the difference of $\hat{T}$ and $T$ in the
renormalization of the superficial degree of divergence of (\ref{ind}),
having all subdivergences renormalized according to $\hat{T}$.

{\bf Proof:} We construct the $\Delta_n$ inductively by the following
formula:

\begin{eqnarray}
\tilde{\Delta}_n & := & 
\hat{T}_n \left( \mbox{$\bigotimes$} _{j \in J} f_j \right)
- \sum_{ P \in Part(J) \atop |P|>1} T_{|P|} \Bigl[ \bigotimes_{O_i \in P}
\Delta_{|O_i|} \left( \mbox{$\bigotimes$}_{j \in O_i}  f_j \right)
\Bigr] \nonumber \\
& = & 
T_1 \left( \Delta_n \left( \mbox{$\bigotimes$} _{j \in J} f_j 
\right)\right). \label{poi}
\end{eqnarray}

To show that this construction makes sense, we prove by induction
over $n$
that the support of $\tilde{\Delta}_n$ is contained in $D_n$.
Since $T_1$ is surjective, there is a $\Delta_n \left( \bigotimes_j
f_j \right) \in \mathcal{D} (\R ^4, \mathcal{A})$ with $T_1 (
\Delta_n ) = \tilde{ \Delta}_n$.
Equation (\ref{msatz}) is automatically fulfilled by this
construction. 
Now we show by induction that $\mathrm{supp} (\tilde{\Delta}_n) \subset D_n$.

Beginning of the induction:

\begin{itemize}
\item $|n|=1$:$ \qquad \tilde{\Delta}_1 (f) = T_1(f)$

\item $|n|=2$:$ \qquad 
\tilde{\Delta}_2 \left( f_1, f_2 \right)  =  \hat{T}_2 
\left( f_1 , f_2
\right) - T_2 \left( \Delta_1 (f_1) \Delta_1 (f_2) 
\right) 
 = \hat{T}_2 \left( f_1 ,f_2 \right)  
- T_2 \left(  f_1,f_2 \right),$

and we saw in (\ref{ind}) that two $T_2$-products are equal
outside the diagonal. So $\mathrm{supp} \tilde{\Delta}_2 \subset D_2$.

\end{itemize}

Now we assume $\cap_{i=1}^n \mathrm{supp} f_i = \emptyset$ and have to show 
$\tilde{\Delta}_n \left( f_1 \ldots f_n \right) =0$.
We obtain with (\ref{ind}):

\begin{eqnarray}
\hat{T}_n  \left( \mbox{$\bigotimes$} _{j \in J} f_j \right)
= \sum _{I \subset J} \sum_k
\hat{T}_{|I^c|} \left( \mbox{$\bigotimes$} _{j \in I^c} 
p_{I,k}^j f_j \right)
\hat{T} _{|I|} \left( \mbox{$\prod$} _{j \in I} 
p_{I,k}^j f_j \right)
\end{eqnarray}

With the induction hypothesis ($\mathrm{supp} \tilde{\Delta} _m \subset D_m$
for all $m<n$) we also obtain a factorization of the second term
of $\tilde{\Delta}_n$

\begin{eqnarray}
\lefteqn{\sum_{ P \in Part(J) \atop |P|>1}
T_{|P|} \Bigl[ \bigotimes_{O_i \in P}
\Delta_{|O_i|} \left( \mbox{$\bigotimes$}_{j \in O_i} f_j) \right)
\Bigr] =} & &  \nonumber \\
& = & \sum_{I \subset J} \sum_k
\sum_{ S \in Part(I^c) \atop T \in Part(I)}
T_{|S|} \Bigl[ \bigotimes_{O_i \in S}
\Delta_{|O_i|} \left( \mbox{$\bigotimes$}_{j \in O_i} 
p_{I,k}^j f_j \right) \Bigr]
T_{|T|} \Bigl[ \bigotimes_{U_i \in T}
\Delta_{|U_i|} \left( \mbox{$\bigotimes$}_{j \in U_i} 
p_{I,k}^j f_j \right) \Bigr]
\end{eqnarray}

(since the support of the $\Delta_{|O_i|}$ is contained in a set
of points belonging to a partition $I \subset J$ with $O_i \subset I$
or $O_i \subset I^c$). Therefore we obtain

\begin{eqnarray}
\lefteqn{ 
\tilde{\Delta}_{|J|} \left( \mbox{$\bigotimes$}_{j \in J} f_j \right)
=} & &  \nonumber \\
& = & \sum_{I \subset J} \sum_k \Biggl\{
\hat{T} _{|I^c|}
\left( \mbox{$\bigotimes$}_{j \in I^c} p_{I,k}^j f_j
\right) \left( 
\hat{T}_{|I|} \left( \mbox{$\bigotimes$}_{j \in I} 
p_{I,k}^j f_j) \right)
- \sum_{ P \in Part(I)} T_{|P|} \Bigl[ \bigotimes_{O_i \in P}
\Delta_{|O_i|} \left( \mbox{$\bigotimes$}_{j \in O_i} 
p_{I,k}^j f_j \right) \Bigr] \right) \nonumber \\
& + &  \left( 
\hat{T}_{|I^c|} \left( \mbox{$\bigotimes$}_{j \in I^c} 
p_{I,k}^j f_j \right)
- \sum_{ P \in Part(I^c)} T_{|P|} \Bigl[ \bigotimes_{U_i \in P}
\Delta_{|U_i|} \left( \mbox{$\bigotimes$}_{j \in U_i} 
p_{I,k}^j f_j \right) \Bigr] \right) \cdot \nonumber \\
& & \cdot \sum_{ P \in Part(I)} T_{|P|} \Bigl[ \bigotimes_{O_i \in P}
\Delta_{|O_i|} \left( \mbox{$\bigotimes$}_{j \in O_i} 
p_{I,k}^j f_j \right)
\Bigr] \Biggr\} = 0 
\end{eqnarray}

because the $\Delta_n$ of lower order fulfill equation (\ref{msatz})
$\blacksquare$.

\section{Renormalization as an Extension of Distributions} \label{rendis}

\subsection{Motivation}

In this section we show that the extension of numerical distributions
corresponds to renormalization.
We derive some properties of our extension procedure,
compare it with the usual way of renormalization in momentum
space and apply it explicitly in the next section
to the renormalization of the second and third
order of the $S$-matrix in $\Phi ^4$-theory, namely:

\begin{eqnarray}
S^{\left( 2 \right)} \left( g \right) & = & \frac{1}{2} \left(
\frac{i\lambda}{4!} \right)^2 
T_2 \left( g(x_1) \phi^4, g(x_2) \phi^4  
 \right) = \frac{1}{2} \left(
\frac{i\lambda}{4!} \right)^2 
T_2 \left( (g \phi^4)^{\otimes 2}\right) \\
S^{\left( 3 \right)} \left( g \right) & = &- \frac{1}{3!} \left(
\frac{i\lambda}{4!} \right)^3  
T_3 \left( g(x_1) \phi^4 ,  g(x_2) \phi^4 ,  g(x_3) \phi^4 
\right) = - \frac{1}{3!} \left(
\frac{i\lambda}{4!} \right)^3 T_3 \left(  
(g \phi^4)^{\otimes 3}\right).
\end{eqnarray}

By means of Wick's theorem the $T_n$-products appearing in the expansion
of the $S$-matrix can be transformed into integrals of
linear combinations of 
products of numerical distributions with Wick products.
For instance $T_2$ in $S^{\left(
2 \right)} \left( g \right)$ has the form:

\begin{eqnarray}
T^0_2 \left( g(x_1) \phi^4 g(x_2) \phi^4 
 \right) & = & \sum_{k=0}^4 
\left( \begin{array}{c} 4 \\ k  \end{array}\right)
\left( \begin{array}{c} 4 \\ k  \end{array}\right)
\left( 4-k\right)! \cdot \nonumber \\
& & \cdot \int d^4 x_1 \int d^4 x_2 \  
t^0(x_1, x_2) 
:\phi^k \left( x_1 \right)
\phi^k \left( x_2 \right): g(x_1) g(x_2) . \label{s} \\
\mbox{with} \qquad \qquad
t^0 (x_1, x_2) & \stackrel{x_1 \not= x_2}{=}& 
\left( i \Delta _F \left( x_1 - x_2 \right) \right)^{4-k}.\nonumber 
\end{eqnarray}

From theorem 0 of \cite{epstein} we know that the product of a welldefined
translation invariant
numerical distribution with a Wick product is welldefined on $D_0$. 
Therefore we only have to take care of 
the numerical distributions, denoted with $t^0$ in the
following. 
In renormalization theory, $t^0$ is for noncoincident points
a product of Feynman propagators, therefore
it is Poincar\'{e} invariant in Minkowski space.
Problems arise from the fact that
Feynman propagators are distributions and the product of 
distributions is not always defined. The domain of definition depends
on the singularities of the individual factors,  
characterized by methods of the microlocal analysis.

\subsection{Some Microlocal Analysis}

The content of this subsection can be found in \cite{123} \cite{hoer} 
\cite{fred}.
Let $\mathcal{M}=\R^{4n}$ or $\mathcal{M} \subset \R^{4n}$
be a manifold of dimension $4n$ and $\mathcal{D}(\mathcal{M})$
all $\mathcal{C}^{\infty}$-functions on $\mathcal{M}$ with 
compact support.
The singular support of a distribution $u \in \mathcal{D}'(\mathcal{M})$,
$sing \ \mathrm{supp} \ u$, is defined as the set of all points in $\mathcal{M}$ 
without any open neighbourhood
to which the restriction of $u$ is a $\mathcal{C}^{\infty}$ function.

If a distribution $v$ with compact support has no singularities, its
Fourier transform $\hat{v}$ is asymptotically bounded
for large $\xi$ by

\begin{eqnarray}
| \hat{v} \left( \xi \right) | \le C_N \left( 1 + |\xi | \right)^{-N} 
\qquad \qquad \forall \ N \in \mathbb{N}, \xi \in \R ^{4n}, \label{bed}
\end{eqnarray}
where $C_N$ are constants for each $N$.
Any distribution $u \in \mathcal{D}'(\mathcal{M})$ is regular in 
$Y \subset \mathcal{M}$, if for all functions $f \in \mathcal{C}^{\infty}_0 (Y)$
the Fourier transform of $fu$ is asymptotically bounded 
for large $\xi$ by

\begin{eqnarray}
| \widehat{fu} \left( \xi \right) | \le C_N \left( 1 + |\xi | \right)^{-N} 
\qquad \qquad \forall \ N \in \mathbb{N}, \xi \in \R ^{4n}, \label{oha}
\end{eqnarray}

where $C_N$ are constants for each $N$.
With $sing \ \mathrm{supp} \ u$ we denote the set of points of 
$\mathcal{M}$ where $u$ is not regular.

Let $u$ be singular at $x \in X \subset \mathcal{M}$, 
$x \in sing \ \mathrm{supp} \ u$.
$\Sigma _x \subset \R^{4n} \setminus 0$ is the set of all $ \xi$
of the cotangent space $T_x^*(X)$ such that (\ref{oha}) is not fulfilled for
any function $f \in \mathcal{C}^{\infty}_0 (X)$ with $f (x) \not=
0$ in a conic neighbourhood V of $\xi$.
$\Sigma _x$ is a cone describing the direction of the high frequencies
causing the singularities at $x$. The pair of $x$ and $\Sigma _x$
is an element of the wave front set:

\begin{Def}

If $u \in \mathcal{D}' \left( \mathcal{M} \right)$, then the closed subset
of $\mathcal{M} \times \left( \R^{4n} \setminus \{0 \} \right) 
\subset T^*(\mathcal{M})$, defined by

\begin{eqnarray}
WF \left(u \right) = \left\{ (x,\xi) \in \mathcal{M} \times 
\left( \R^{4n} \setminus \{0\} \right)
 \ | \ \xi \in \Sigma _x \left( u \right) \right\}
\end{eqnarray}

is called the wave front set of $u$. The projection on $\mathcal{M}$ is
$sing \ \mathrm{supp} \ u$.
\end{Def}

Multiplication with a smooth function $a \in \mathcal{C}^{\infty}$
and differentiation do not enlarge the wave front set:

\begin{eqnarray}
WF \left( au \right) & \subset& WF \left( u \right), \\
WF \left( D^{\alpha}u \right) & \subset& WF \left( u \right).
\end{eqnarray}

For the existence of the pointwise product of two 
distributions $u,v$ at $x$ it is sufficient to fulfill the condition

\begin{eqnarray}
(x,0) \notin  WF \left( u \right) \oplus WF \left( v \right)=
\left\{ (x, \xi _1 + \xi _2)| (x,\xi _1) \in  WF \left( u \right),
 (x,\xi _2) \in  WF \left( v \right) \right\}.
\end{eqnarray}

Thus the product at $x$ exists, if $u$ or $v$ or both are regular
in $x$. If  $u$ and $v$ are singular at $x$, the product exists,
if the second components in the wave front sets of $u$ and $v$ at $x$
cannot be added to the zerovector.
If the product exists, its wave front set will fulfill

\begin{eqnarray}
WF \left( uv \right) \subset WF \left( u \right) \cup
WF \left( v \right) \cup \left( WF \left( u \right) \oplus
WF \left( v \right)  \right). \label{wfbed}
\end{eqnarray}

We now check if the products of Feynman propagators of scalar fields
appearing in (\ref{s}),

\begin{eqnarray}
\left( i \Delta _F \left( x_1 - x_2
\right) \right)^n  \qquad \mbox{with} \ \ n= 2,3,4 \label{produkt}
\end{eqnarray}

exist for all $ x_1 - x_2$.
Since $ i \Delta _F \left( x_1 - x_2 \right)$ is for $x_1 \not= 
x_2$ a solution
of the Klein-Gordon equation, its singular support is contained in
the characteristic set of the Klein-Gordon operator, the
forward and backward lightcone. 

The wave front set set of the Feynman propagator has the following form:

\begin{eqnarray}
WF \left( \Delta _F \right) & = & 
\{ (x_1,k; x_2, k') \in T_{x_1}^* \R^4
\times \left. T_{x_2}^* \R^4 \right| (x_1,k) \sim (x_2,-k'), k \in \bar{V}
_{\stackrel{+}{-}} 
\mbox{if} \ x_1 \in J_{\stackrel{+}{-}} (x_2) \} \nonumber \\
& & \cup \{ (x_1, k; x_1, -k), k \not= 0\}.
\end{eqnarray}

Here $\bar{V}_{\stackrel{+}{-}}$ is the closed forward respectively backward
light cone and $J_{\stackrel{+}{-}} (x_2)$ are all points in 
$\mathcal{M}$ in the future respectively past of $x_2$
which can be connected with $x_2$ by a causal curve $\gamma$.
$(x_1,k) \sim (x_2,m)$ means that $x_1$ and $x_2$ can be connected
by a light cone with cotangent vectors $k$ and $m$ at $x_1$ and $x_2$.

The sum of the second components of the wave front sets
of two propagators can
vanish only at $x_1 = x_2$.
Therefore the products (\ref{produkt}) 
are defined on $\left( \R^4 \right)^2 \setminus D_2$.

With (\ref{wfbed}) we are also able to investigate higher products of
Feynman propagators occuring in higher orders of the $S$-matrix.
For instance we discuss the wave front set of the numerical distribution

\begin{eqnarray}
t = \underbrace{\left( i \Delta _F (x_1 - x_2) \right)^2}_{u} 
\underbrace{\left( i \Delta _F (x_2 - x_3) \right)^2}_{v} 
\underbrace{\left( i \Delta _F (x_1 - x_3) \right)}_{w}. \label{dist}
\end{eqnarray}

which occurs in our third order calculation of the $S$-matrix in
$\phi^4$-theory.

\begin{eqnarray}
WF(uvw) & \subset & WF(u) \cup WF(v) \cup WF(w) \nonumber \\
& & \cup \left( WF(u) \oplus WF(v) \right) \cup
\left( WF(w) \oplus WF(v) \right) \cup
\left( WF(u) \oplus WF(w) \right) \nonumber \\
& & \cup \left( WF(u) \oplus WF(v) \oplus WF(w) \right).\label{wellen}
\end{eqnarray}

The wave front sets in the first line of (\ref{wellen}) yield problems
for $x_3 \not= x_2 = x_1$ (in $WF(u)$) and for $x_2 = x_3 \not= x_1$
(in $WF(v)$), the ill-definedness of the product on this subset
is treated in the renormalization of subdivergences.
The remaining sums of wave front sets can contain a zero component for
$x_1 = x_2 = x_3$. This ill-definedness of the numerical distribution
is called the superficial divergence.

\subsection{Power Counting of Divergences}

In momentum space calculations the ill definedness of products
of Feynman propagators corresponds to UV divergences of loop
integrals. Divergent terms can be found by counting
the powers of momenta in the loop integrals (power counting).
The superficial degree of divergence of a diagram corresponds
in position space to the singular order at the total diagonal
of the numerical distribution belonging to this diagram. We first
introduce the definition of the scaling degree of a distribution:

\begin{Def}
The scaling degree of a numerical distribution $t \in 
\mathcal{D}' (\R^{4n})$ at the origin is defined by

\begin{eqnarray}
sd(t) := \inf \left\{ \omega  \left| \lim_{\epsilon \to 0}
\epsilon^{\omega} t \left( \epsilon x_1 , \ldots, 
\epsilon x_n \right) =0 \ \mbox{in the sense of
distributions} \right. \right\}. \label{scaling}
\end{eqnarray}

\end{Def}

The singular order is now given by

\begin{Def}
The singular order of a numerical distribution $t \in \mathcal{D}' 
(\R^{nd})$
at the origin is defined by

\begin{eqnarray}
sing \ ord \ t = [sd(t)] - nd \label{sing}
\end{eqnarray}

where $d$ is the space time dimension.
\end{Def}

There exist different definitions of the scaling degree in the literature.
The definition above is the Steinmann scaling degree \cite{steinmann}.
The singular order of a distribution with respect to this scaling degree
is not larger than the superficial degree of divergence obtained by
power counting in momentum space \cite{fredneu}.
In \cite{fredneu} is also given a definition of a scaling degree
of a distribution with respect to
a submanifold, called microlocal scaling degree.
With this microlocal scaling degree we determine all scaling degrees
of the numerical distributions in our calculations, especially we
obtain

\begin{eqnarray}
sd \left( \left( i \Delta _F \left( x_1 - x_2\right)\right)^{4-k}  
\right) & = & 8-2k,
\end{eqnarray}

which can be also obtained by using (\ref{scaling}).

At $x_1 = x_2$ the
products of Feynman propagators have the singular order

\begin{eqnarray}
sing \ ord \left( \left( i \Delta _F \left( x_1 - x_2\right)\right)^{4-k}  
\right) = 4 - 2k. \label{fsingord}
\end{eqnarray}

For the distribution t in (\ref{dist}) we obtain
$sd (t) = 10$ at $x_1 = x_2 =x_3$ whereas at $x_1 = x_2 \not= x_3$
and at $x_2 = x_3 \not= x_1$ we have $sd(t) = 4$ and at
$x_2 \not= x_1 = x_3$ it holds $sd (t) = 2$.

The singular order of the superficial divergence is 2, the
singular orders corresponding to
the subdivergences are 0 and -2.

\subsection{Extension of Distributions}

In momentum space integrals of diagrams with negative degree of
divergence are finite in Euclidean calculations. To give also divergent
diagrams a sense one has to renormalize them, and this procedure
is not unique. Correspondingly we have the following two theorems
in position space, which are proved in \cite{fredneu} on curved
space times. The first one states that
the extension of a distribution of negative singular order exists
and is unique. The proof can be already found in \cite{fred}:

\begin{Thm} \label{th1}

If $t^0 \left( x_1, \ldots ,x_n \right) \in \mathcal{D}' 
\left( \R^{4n} \setminus 0 \right)$
has singular order $\delta < 0$ at the origin, then a
unique $t \left( x_1, \ldots ,x_n \right) \in \mathcal{D}' 
\left( \R^{4n}  \right)$ exists with the same singular order at
$0$ and

\begin{eqnarray}
t^0 \left( \phi \right) = t \left( \phi \right) \qquad \qquad
\forall \ \phi \in \mathcal{D} 
\left( \R^{4n} \setminus 0 \right).
\end{eqnarray}

\end{Thm}

Distributions with zero or positive singular order $\delta$
can be extended, but the extension is not unique.
We first remark that they are only defined on test functions
vanishing sufficiently fast at $0$:

\begin{eqnarray}
\infty > sing \ ord \ t^0 = \delta \geq 0
\qquad \Rightarrow \qquad 
t^0 \in \mathcal{D}'_{\delta} 
\left( \R^{4n}  \right)
\end{eqnarray}
 
with

\begin{eqnarray}
\mathcal{D}_{\delta} 
\left( \R^{4n} \right)
& = & \left\{ \phi \in \mathcal{D} (\R^{4n}) \ | \ D^{\alpha} \phi
\left( 0 \right) =0 \quad
\mbox{for all multiindices } \right. \nonumber \\ 
& & \quad  \alpha = \left.
\left( \alpha_1, \ldots, \alpha_n \right) \ \mbox{with} \
|\alpha| \leq 
\delta \right\}.
\end{eqnarray}

\begin{Thm} \label{th2}

For all 
$t^0 \left( x_1, \ldots ,x_n \right) \in \mathcal{D}'_{\delta} 
\left( \R^{4n} \right)$
with $sing \ ord \ t^0 = \delta$, $0 \leq \delta < \infty$,
at $0$ exist numerical distributions
$t \in \mathcal{D}' \left( \R^{4n} \right)$ with the
same singular order $\delta$ at $0$ and

\begin{eqnarray}
t^0 \left( \phi \right) = t \left( \phi \right) \qquad \qquad
\forall \ \phi \in \mathcal{D}_{\delta} 
\left( \R^{4n} \right).\label{th2for}
\end{eqnarray}

\end{Thm}

In \cite{fredneu} this is proved on curved space times.

To construct extensions of distributions with positive
singular order, a projection operator on testfunctions is used.

\begin{Def}

With

\begin{eqnarray}
W^{(k)}\left( \delta, w, (x_1, \ldots,x_k)\right): 
\ \mathcal{D} \left( \R^{4n} \right) \longrightarrow
 \ \mathcal{D}_{\delta}  \left( \R^{4n} 
\right)  \nonumber 
\end{eqnarray}

\begin{eqnarray}
\phi \left( x_1, \ldots,x_n \right) & \mapsto & 
\phi \left( x_1, \ldots,x_n \right) - \nonumber \\
& & -w (x_1) \cdot \ldots \cdot w(x_k) 
\sum^{|\alpha| \leq \delta} _{\alpha_i=0 \ \forall i>k}
\frac{x^{\alpha}}{\alpha !}
D^{\alpha} \phi \left( x_1, \ldots,x_n \right) 
\Bigr| _{x_1 = \ldots =x_k =0} \nonumber 
\end{eqnarray}

we define a projection operator on testfunctions of the order $\delta$ in 
$x_1 \ldots x_k$
for each function $  w  \in  \mathcal{C} _0^{\infty} 
\left( \R^4 \right)$ fulfilling $w(0)=1$ and $D^{\alpha}  w(0) =0 $
for all multiindices $\alpha$ with $ 0< |\alpha| \leq \delta$ and
$\alpha_i=0$ for all $i>k$.

\end{Def}

$W^{(k)}$ is a modified Taylor subtraction operator 
in $k$ variables of the testfunctions.
Since the function $w$ has compact support, the result is a function with 
compact support, and vanishes up to the order $\delta$ at
$0$. Therefore the singular distribution
$t^0 \left( x \right) $ with singular order $\delta$
is defined on all
$ W^{(n)}\left( \delta, w, x \right) 
\phi \left( x \right)$ with $x=(x_1, \ldots x_n)$.
For $k<n$ the operator $W^{(k)}$ is used in the renormalization
of subdivergences.

The following construction of the ("superficial")
extension of a numerical
distribution $t^0$, welldefined outside the origin,
with $sing \ ord \ t^0 = \delta$
fulfills (\ref{th2for}):

\begin{eqnarray}
< t (x),\phi (x)> & = & < t^0(x), W^{(n)}\left( 
\delta, w, x \right)
\phi (x)> + \nonumber \\
& & + \sum_{|\alpha| \leq \delta} 
\frac{(-1)^{|\alpha|}c^{\alpha}}{\alpha !}
<\delta^{\alpha} ( x ), \phi ( x )>.  \label{fort}
\end{eqnarray}

The free constants $c^{\alpha}$ express the ambiguity in the
extension of the distribution, which remains after the function
$w(x)$ in the $W$-operator is fixed,
it holds

\begin{eqnarray}
<t,w x^{\alpha}> = c^{\alpha} \qquad
\mbox{for} \ |\alpha| \leq \delta.
\end{eqnarray}

The form of the $c^{\alpha}$ can be further restricted by demanding
the invariance of the distribution $t$ under symmetry operations,
the Lorentz invariance for instance.
The most simple choice of the  $c^{\alpha}$ would be $c^{\alpha}=0$.
But we will see that this choice is incompatible with Lorentz
invariance for $ |\alpha|>1$.
Another choice of the $c^{\alpha}$ leads to the form of
BPHZ subtraction at momentum $q$ in momentum space. The explicit form of
these $c^{\alpha}$ is given in \cite{scharf}. Naturally this choice of the
$c^{\alpha}$ for $q \not= 0$ is again incompatible with
the Lorentz invariance for $ |\alpha|>1$.

In the following we will restrict ourselves to the treatment of extensions
of the form (\ref{fort}).

\subsection{The Lorentz Invariant Extension in Scalar Theories}

It is possible to determine the free constants $c^{\alpha}$
of an extension, so that the result is a
Lorentz invariant distribution.
In \cite{wir} the main idea of the calculation is introduced and
an explicit result is obtained for scalar distributions in one variable.
A further development of the techniques yields an inductive
formula for a Lorentz invariant extension of an arbitrary distribution
\cite{harakiri}. In our calculations of the second and third order 
of the $S$-matrix in $\Phi ^4$-theory we only need
the results of \cite{wir}, which are repeated here.
We use the notation

\begin{eqnarray}
b^{(1 \ldots n)}= \frac{1}{n!} \sum_{\pi \in S_n} b^{\pi (1)
\ldots \pi (n)}
\end{eqnarray}

for the total symmetric part of the tensor $b$.
With the following abbreviations for $n \in \N$,

\begin{eqnarray}
n!! = \left \{ \begin{array}{r@{\quad}l}
2 \cdot 4 \cdot \ldots \cdot n & \mbox{ for $n$  even} \\
1 \cdot 3 \cdot \ldots \cdot n & \mbox{ for $n$  odd}
\end{array} \right.
\end{eqnarray}

and

\begin{eqnarray}
\left[ \frac{n}{2} \right] =
\left \{ \begin{array}{r@{\quad}l}
\frac{n}{2} & \mbox{ for $n$  even} \\
& \\
\frac{n-1}{2} & \mbox{ for $n$  odd}
\end{array} \right.
\end{eqnarray}

and $x^2 = x^{\mu}x_{\mu}$,
the symmetric part of $c^{\alpha}$ is given by

\begin{multline}
  c^{(\alpha _1 \ldots \alpha_n)} = \frac{ (n-1)!!}{(n+2)!!}
  \sum_{s=0}^{\left[\frac{n-1}{2} \right] }
  \frac{(n-2s)!!}{(n-1-2s)!!}  g^{(\alpha_1 \alpha _2} \ldots
  g^{\alpha _{2s-1} \alpha _{2s}} \times
  \\ \times
  \left\langle {t^0}, (x^2)^s x ^{\alpha _{2s+1}} \ldots x ^{\alpha
      _{n-1}} \left( x^2 \partial ^{\alpha _{n})} w - x ^{\alpha
        _{n})} x^{\beta} \partial _{\beta} w \right) \right\rangle.
\label{result}
\end{multline}

This result is unique up to Lorentz invariant terms. 
Choosing the coefficients according to (\ref{result}) the remaining freedom
in the renormalization procedure is the choice of the function $w$
and these Lorentz-invariant counterterms.
In the following we set all counterterms not depending on $w$ 
equal to 0, so only the contributions of (\ref{result}) are
nonvanishing.
In the 
calculations of the next section we need the coefficients $c^{(\alpha)}$
for $|\alpha|=0$ and $|\alpha|=2$. In the case $|\alpha|=0$ all 
choices of $c$ are Lorentz invariant, and we set $c=0$.
For $|\alpha|=2$ we have

\begin{eqnarray}
c^{(\alpha _1 \alpha_2)} = - \frac{1}{4} \left\langle t^0,  x ^{\alpha_1}
x ^{\alpha_2} x ^{\sigma} \partial _{\sigma}w - x^2 x^{(\alpha_1}
\partial ^{\alpha_2 )}w \right\rangle.
\end{eqnarray}

In $\Phi ^4$-theory we want to renormalize diagrams with 2 and
4 external legs. Therefore we have $sing \ ord \ t \leq 2$, and in this
case we obtain by partial integration

\begin{eqnarray}
c^{(\alpha _1 \alpha _2)} & = & - \frac{1}{4}
\left\langle \partial ^{(\alpha _2} x^2 x^{\alpha _1)} t^0
- \partial _{\beta} x^{\alpha _1}  x^{\alpha _2}
x^{\beta} t^0, w \right\rangle.
\end{eqnarray}

The numerical distributions are products of Feynman propagators
and depend only on $x^2$: 

\begin{eqnarray}
\partial_{\sigma} t^0 = 2 x_{\sigma} (t^0)'
\end{eqnarray}

Therefore we have

\begin{eqnarray}
c^{\alpha _1 \alpha _2} & = & 
\left\langle x^{\alpha _1}  x^{\alpha _2} t^0, w \right\rangle
- \frac{1}{4} \left\langle x^2 g^{\alpha _1 \alpha _2} t^0, 
w \right\rangle \nonumber \\
& = & \left\langle t^0, (  x^{\alpha _1}  x^{\alpha _2}
- \frac{1}{4} x^2 g^{\alpha _1 \alpha _2})w \right\rangle.
\end{eqnarray}

In the following calculations only terms even in the variables
contribute. Since only the support of the function
$w$ is important, we can assume without loss of generality
that the function is even, too.
It follows that the Lorentz invariant extension of a
distribution of singular order 2 according to (\ref{fort}) has the
form

\begin{eqnarray}
\left\langle t, \phi \right\rangle & = & \left\langle t^0,
\phi(x) - w(x) \phi (0) - \frac{w(x)}{8} x^2 \Box \phi (0)
\right\rangle. \label{fortform}
\end{eqnarray}

\subsection{Properties of the $W$-Operator} \label{wop}

\begin{enumerate}

\item As a projection operator, $ W^{(k)}$ has the following properties:

$W^{(k)2} = W^{(k)}$.

For $n \geq l \geq m$, $\delta \geq \delta'$, the following
relation holds:

\begin{eqnarray}
W^{(l)}\left( \delta', w', (x_1, \ldots,x_l)\right) 
W^{(m)}\left( \delta, w, (x_1, \ldots,x_m)\right) 
\phi \left( x_1, \ldots,x_n \right) = \nonumber \\ 
W^{(m)}\left( \delta, w, (x_1, \ldots,x_m)\right) 
\phi \left( x_1, \ldots,x_n \right).
\end{eqnarray}

\item The renormalization scale

Characteristic for every renormalization procedure is the
occurence of a mass scale, the renormalization scale.
In the extension of the $T$-products the $W$-operation
depends on a function $w(x)$.
Since the argument of this function should be dimensionless,
it depends implicitly on a mass scale: $w= w(mx)$.
The shape of the function $w$ describes the subtraction procedure,
in the region of $w=1$ the subtraction is the full Taylor subtraction,
whereas nothing is subtracted outside the support of $w$.

Varying the mass scale $m$ with fixed $w$, we change the support
region of the function $w$ and regulate in this way the subtraction
procedure.
In the limit $m \to 0$ we have $w(mx) \equiv 1$ and the
Taylor subtraction acts everywhere, whereas in the limit
$m \to \infty$ the support of $w(mx)$ shrinks to the origin
and we subtract only at this point.

In the following, we continue to write $w(x)$, and only if we
need the dependence on the mass scale, we will write this function as 
$w(mx)$. 

A variation of the mass scale yields

\begin{eqnarray}
w((m+ \delta m)z)- w(mz) = \delta m \frac{\partial}{\partial m}
 w(mz) =   \delta m  z^{\mu} \partial_{\mu}  w \label{xyz1}.
\end{eqnarray}

\end{enumerate}

\section{Renormalization of the $S$-Matrix in $\Phi ^4$-Theory}

\subsection{Introduction}

In this section the second and third order of the $S$-matrix
in  $\Phi ^4$-theory are renormalized using the formalism
developped in the previous section. Already in second order we
will see that the operators $\tilde{\Delta}_n$ of theorem
(\ref{meinsatz}) consist only of linear combinations of
$:\partial _{\mu} \phi \partial ^{\mu} \phi:, :\phi ^2:$ and
$:\phi ^4:$. In third order we demonstrate an explicit
calculation of a diagram with subdivergences to make clear
how the subtraction works and that the result is
indeed independent of the partition of unity used in the inductive
construction. 
The result of the calculations is given in the adiabatic limit,
otherwise there would be many additional terms.

Before starting the calculations it is useful to
make some remarks.
We will have to work with expressions of the typical form

\begin{eqnarray}
\int d u
\int d x \ t^0 (u) \ 
\left( W \left( \delta, w, u \right)
A (x, u) g(x,u) \right),
\end{eqnarray}

where $t^0(u)$ is the numerical distribution with singular order
$\delta$, $g$ is a testfunction with compact support
and $A$ is a Wick product of fields at $x$ and $u$.
With $u$ and $x$ we denote a tupel of coordinates $(u_1, \ldots u_n)$
resp. $(x_1, \ldots x_n)$.
We always assume that the function $w(u)$ of the $W$-operation 
is even in all components of $u^{\mu}_i$ for all $i$.
In the calculations we use the following facts:

\begin{enumerate}

\item
If the distributions $t^0$ are even in all $u^{\mu}_i$, all
odd terms in the Taylor
subtraction of $W$ will vanish due to the integration over $u$.

\item
To make the extension Lorentz invariant, we use the form (\ref{result}) 
respectively (\ref{fortform}) of the subtraction.
Here again the contributions of the coefficients $c^{\alpha}$
with $|\alpha|$ odd vanish, if $t^0$ is even in the $u^{\mu}_i$.

\item
For $\delta \geq 1$ there appear terms with derivatives of $g$ in
the Taylor subtraction. These terms vanish in the adiabatic limit.
Because the limit exists in massive $\Phi ^4$-theory in the
strong sense, we omit them from the beginning.

\item
In \cite{michael} it is shown that the adiabatic limit of 
vacuum diagrams only exists for a special choice of the $c^{\alpha}$.
This choice is Lorentz invariant, and
the contributions of the vacuum diagrams vanish in the 
adiabatic limit, too.

\end{enumerate}

We denote with $()_R$ the extension
of a numerical distribution and with $()_E$ the extension of a
$T_m$-product to the total diagonal $D_m$.

\subsection{Renormalization of the Second Order} \label{2ord}

The second order term in the $S$-matrix has the form

\begin{eqnarray}
S^{\left( 2 \right)} \left( g \right) & = & \frac{1}{2} \left(
\frac{-i\lambda}{4!} \right)^2  
T_2 \left( g(x_1) \Phi^4, g(x_2) \Phi^4  \right)  \nonumber \\
& = & - \frac{\lambda ^2}{2 (4!)^2} \int d^4 x_1 \int d^4 x_2 
\sum_{k=0}^4 
\left( \begin{array}{c} 4 \\ k  \end{array}\right)
\left( \begin{array}{c} 4 \\ k  \end{array}\right)
\left( 4-k \right)! \ \cdot \nonumber \\
& & \cdot \left( \left( i \Delta _F \left( x_1 - x_2
\right) \right)^{4-k}\right)_R 
 :\phi^k \left( x_1 \right)
\phi^k \left( x_2 \right):
 g \left( x_1 \right)  g \left( x_2 \right). \label{szwei}
\end{eqnarray}

With the singular orders  
(\ref{fsingord}) of the Feynman propagators we obtain
nontrivial contributions of the extension 
from terms with $k=0,1,2$.

\begin{enumerate}

\item $k=0$ yields a vacuum diagram. Its contribution vanishes
in the adiabatic limit.

\item For $k=1$ we have

\begin{eqnarray}
S_{(k=1)}^{\left( 2 \right)} \left( g \right)  & = &
- \frac{96 \lambda ^2}{2!(4!)^2}  \int d^4 x_1 \int d^4 x_2 
\left( i \Delta _F \left( x_1 - x_2
\right) \right)^3_R \cdot \nonumber \\
& & \qquad \qquad \qquad \cdot 
 :\phi \left( x_1 \right) \phi \left( x_2 \right):
 g \left( x_1 \right)  g \left( x_2 \right) \nonumber \\
& \stackrel{(\ref{fortform})}{=}&
- \frac{\lambda ^2}{12}  \int d^4 v  \int d^4 u 
\left( i \Delta _F \left( u \right) \right)^3 \Bigl[
:\phi (u+v) \phi (v): g(u+v) g(v) \nonumber \\
& & 
- w(u) :\phi ^2 (v): g^2 (v) 
- \frac{1}{8} w(u) u^2 :\phi (v) \Box \phi (v):
g^2 (v) \Bigr].
\end{eqnarray}

\item For $k=2$ we obtain

\begin{eqnarray}
S_{(k=2)}^{\left( 2 \right)} \left( g \right)  & = &
- \frac{\lambda ^2}{16} \int d^4 u \int d^4 v 
\left( i \Delta _F \left( u \right) \right)^2 \nonumber \\
& & \Bigl[ : \phi ^2 (u+v) \phi ^2 (v): g(u+v) g(v)
- w(u) :\phi ^4 (v): g^2 (v) \Bigr]. \nonumber \\
\end{eqnarray}

\end{enumerate}

{\bf Remark}: The individual terms in the square brackets
are not defined, only their sum is convergent and 
welldefined. The result depends on the function $w(x)$. We
use this dependence to read off the form of the finite
renormalizations $\tilde{\Delta}_2$:
if we choose another Lorentz -invariant $T$-product with
extensions of the form (\ref{fort})
in which all nonvanishing coefficients $c$ depend on $w(x)$,
only the function $w(x)$ 
will differ. The difference of two $w$-functions
has no support at 0 and gives a welldefined contribution.
Denoting with $\hat{T}$ the renormalization at the scale $m + \delta m$
and with $T$ the renormalization at $m$ we obtain with (\ref{xyz1}):

\begin{eqnarray}
\tilde{\Delta}_2 & = &
\left( S_{\hat{T}}^{(2)}- S_T^{(2)} \right)  =  
\nonumber \\ 
& = & 
\int d^4 v \left[ A^{(2)}
:\phi ^2 (v):  + B^{(2)}  
:\partial_{\mu} \phi (v) \partial^{\mu} \phi (v): 
+ C^{(2)} :\phi ^4 (v):  \right] g^2 (v) 
\end{eqnarray}

with

\begin{eqnarray}
A^{(2)} & = & \frac{\lambda ^2}{12} \delta m \
\int d^4 u  
\left( i \Delta _F (u) \right)^3 
\  u^{\mu} (\partial_{\mu}  w) 
\nonumber \\
B^{(2)} & = & - \frac{\lambda ^2}{96} \delta m \
\int d^4 u  
\left( i \Delta _F (u) \right)^3 
\ u^{\nu} (\partial_{\nu}  w) u^2
\nonumber \\
C^{(2)} & = & + \frac{\lambda ^2}{16} \delta m \
\int d^4 u  
\left( i \Delta _F (u) \right)^2 
\ u^{\mu} (\partial_{\mu}  w).
\end{eqnarray}

At this stage we are able to see how the theoretically predicted form of 
the higher $\tilde{\Delta}_n$ is realized in the calculations. 
$\tilde{\Delta}_n$ is the difference in the superficial renormalization 
of diagrams with four and two external legs. Diagrams with four external
legs are superficial logarithmic divergent and of the form

\begin{eqnarray}
\int du \int dv \  t^0 (v,u) W(0,w,u) : \phi \left( l_1(u,v) \right)
\phi \left( l_2(u,v) \right) \phi \left( l_3(u,v) \right)
\phi \left( l_4(u,v) \right): \nonumber \\
\qquad \qquad \qquad \qquad g\left( l_5(u,v) \right) 
\ldots g\left( l_{n+4}(u,v) \right) 
\end{eqnarray}

where $u=(u_1, \ldots , u_{n-1})$ are the difference variables
and $l_i, i=1, \ldots, n+4$ are linear combinations of $v$ 
and the $u_i$ of the
form $l_i (v,u) = v + a_i u_i$ with $a_i \in \R$.
Varying the mass scale in $W$ we obtain only the following contribution

\begin{eqnarray}
\int dv \ C^{(n)} : \phi ^4 (v): g^n (v).
\end{eqnarray}

Diagrams with two external legs are quadratic divergent and of the form

\begin{eqnarray}
\int du \int dv \ t^0 (v,u) W(2,w,u) : \phi \left( l_1(u,v) \right)
\phi \left( l_2(u,v) \right) :
 g\left( l_3(u,v) \right) 
\ldots g\left( l_{n+2}(u,v) \right). 
\end{eqnarray}

Varying the mass scale in the $W$-operation yields the following
contributions:

\begin{eqnarray}
\int dv \left( A^{(n)} : \phi ^2 (v): + B^{'(n)} : \phi (v) \Box 
\phi (v) \right) g^n (v).
\end{eqnarray}

By partial integration we obtain the contribution

\begin{eqnarray}
\int dv \  B^{(n)} : \partial _{\mu} \phi (v) \partial ^{\mu} \phi (v):
g^n (v)
\end{eqnarray}

because the contributions with derivatives of $g$ vanish in the
adiabatic limit. Finite renormalizations to all orders
in $\Phi ^4$-theory consist only of shifts 
in the coupling constant, in masslike terms and
kinetic terms of $\mathcal{L}_{int}$,
whose coefficients $A^{(n)}$, $B^{(n)}$ and $C^{(n)}$ have to be
determined in the calculations.

\subsection{Renormalization of the Third Order} \label{3ord}

The third order term in the $S$-matrix has the form

\begin{eqnarray}
S_3 (g) & = & \frac{i \lambda ^3}{3! (4!)^3} 
T_3 \left( g(x_1) \phi^4, g(x_2)\phi^4, g(x_3) \phi^4  \right) 
\end{eqnarray}

and with the structure (\ref{ind}) of the higher $T$-products we
obtain

\begin{eqnarray}
S_3 (g)
& = & \frac{i \lambda ^3}{3! (4!)^3} 
\int d^4 x_1 \int d^4 x_2 \int d^4 x_3
\biggl\{
\sum_{l \in \{ 1,2,3\}, m<n \atop  m,n \in 
\{ 1,2,3\} \setminus l}
\nonumber \\
& & \left(  f_l^{(3)} +  f_{mn}^{(3)} \right)
\biggl[ \sum _{i=0}^4 \sum _{j_1 =0}^{4-i} \sum _{j_2=0}^{ min(4-j_1,4-i)}
K (i, j_1, j_2 ) \cdot \nonumber \\
& & \cdot \left( \left( i \Delta _F ( x_m
- x_n ) \right)^{i} \right)_R 
\left( i \Delta _F ( x_l - x_m) \right)^{j_1}
\left( i \Delta _F ( x_l - x_n) \right)^{j_2} \cdot
\nonumber \\
& & \cdot  : \phi ^{4-j_1-j_2}( x_l)  \phi ^{4-i-j_1}( x_m)
\phi ^{4-i-j_2}( x_n): g(x_1) g(x_2) g(x_3) \biggr] \biggr\}_E, \label{ente}
\end{eqnarray}

with the factor

\begin{eqnarray}
K(i, j_1 , j_2) & = & 
\left( \begin{array}{c} 4 \\ i  \end{array}\right)
\left( \begin{array}{c} 4 \\ i  \end{array}\right)
\left( \begin{array}{c} 4 \\ j_1  \end{array}\right)
\left( \begin{array}{c} 4-j_1 \\ j_2  \end{array}\right)
\left( \begin{array}{c} 4-i \\ j_1  \end{array}\right)
\left( \begin{array}{c} 4-i \\ j_2  \end{array}\right)
(i)! j_1! j_2! \nonumber \\
& = & \frac{(4!)^3}{((4-j_1-j_2)! j_1 ! (4-i-j_1)! j_2! (4-i-j_2)!i!}.
\label{77}
\end{eqnarray}

Reordering the terms in expression (\ref{ente}) we arrive at

\begin{eqnarray}
S_3 (g) & = & \frac{i \lambda ^3}{3! (4!)^3} 
\int d^4 x_1 \int d^4 x_2 \int d^4 x_3
\sum _{i=0}^4 \sum _{j_1 =0}^{4-i} \sum _{j_2=0}^{ min(4-j_1,4-i)}
K (i, j_1, j_2 ) \cdot
\nonumber \\
& & \cdot
\biggl\{ (f_{12}^{(3)} + f_3^{(3)})  \left( 
\left( i \Delta_F (x_1-x_2) \right)^{i} \right)_R 
\left( i \Delta_F (x_1-x_3) \right)^{j_1}
\left( i \Delta_F (x_3-x_2) \right)^{j_2}
\nonumber \\
& & 
 +  (f_{13}^{(3)} + f_2^{(3)}) \left( 
\left( i \Delta_F (x_1-x_3) \right)^{j_1} \right)_R 
\left( i \Delta_F (x_1-x_2) \right)^{i}
\left( i \Delta_F (x_3-x_2) \right)^{j_2}
\nonumber \\
& & 
 +   (f_{23}^{(3)} + f_1^{(3)}) \left( 
\left( i \Delta_F (x_2-x_3) \right)^{j_2} \right)_R 
\left( i \Delta_F (x_1-x_3) \right)^{j_1}
\left( i \Delta_F (x_1-x_2) \right)^{i} \biggr\}_R \cdot
\nonumber \\
& & \cdot : \phi ^{4-j_1 - j_2} (x_3) \phi ^{4-i- j_2} (x_2) 
\phi ^{4-i-j_1} (x_1): g(x_1) g(x_2) g(x_3) .\label{78}
\end{eqnarray}

There are two kinds of $W$-operations in the calculation, namely
Taylor subtractions in one and two variables corresponding to
the renormalization of superficial divergences and subdivergences.
In the subtraction in one variable $u$ we use the testfunction
$w(u)$ as before. The testfunction of the $W$-operator in two
variables $u$ and $v$ is chosen to be $\tilde{w}(u)\tilde{w}(v)
\tilde{w}(u+v)$, where $\tilde{w}$ is a testfunction in
one variable. Since $u$ and $v$ are difference variables, 
the function is symmetric in all coordinates.

We now list the 14 topological different diagrams occuring in the
sum with their values of $i, j_1$ and $j_2$. Furthermore 
the singular orders of
the whole diagram $\delta ( x_1, x_2,x_3)$ and of the subdiagrams,
$\delta (x_i, x_j)$ consisting only of the vertices $x_i$ and
$x_j$ are listed. $N$ is the number of all different diagrams 
in the sum with the same topological structure.

\vspace{1cm}

\begin{tabular}{|c||c|c|c||c||c|c|c||c|} \hline
Number & $i$ & $j_1$ & $j_2$ & $\delta ( x_1, x_2,x_3)$ &
$\delta (x_1, x_2)$ & 
$\delta (x_1, x_3)$ & $\delta (x_2, x_3)$ & $N$ \\
\hline \hline
1 & 4 & 0 & 0 & 0 & 4 & -4 & -4 & 3\\ \hline \hline
2 & 3 & 0 & 1 & 0 & 2 & -4 & -2 & 6 \\ \hline
3 & 2 & 2 & 0 & 0 & 0 & 0 & -4  & 3 \\ \hline
4 & 2 & 1 & 1 & 0 & 0 & -2 & -2 & 3   \\ \hline \hline
5 & 3 & 1 & 1 & 2 & 2 & -2 & -2 & 3 \\ \hline
6 & 2 & 2 & 1 & 2 & 0 & 0 & -2  & 3  \\ \hline \hline
7 & 2 & 2 & 2 & 4 & 0 & 0 &  0  & 1 \\ \hline \hline
8 & 3 & 0 & 0 & -2 & 2 & -4 & -4 & 3   \\ \hline
9 & 2 & 1 & 0 & -2 & 0 & -2 & -4 & 6   \\ \hline
10 & 1 & 1 & 1 & -2 & -2 & -2 & -2 & 1   \\ \hline
11 & 2 & 0 & 0 & -4 & 0 & -4 & -4 & 3   \\ \hline
12 & 1 & 1 & 0 & -4 & -2 & -2 & -4  & 3  \\ \hline
13 & 1 & 0 & 0 & -6 & -2 & -4 & -4 & 3   \\ \hline
14 & 0 & 0 & 0 & -8 & -4 & -4 & -4 & 1   \\ \hline
\end{tabular}

\vspace{1cm}

The contributions of the diagrams 1 and 7 vanish in the adiabatic limit.

Diagrams 2, 3 and 4 are superficial logarithmic divergent,
they have four external legs and contribute to the renormalization
of the coupling constant. 
Diagram 5 and 6 have two external lines and are therefore
superficial quartic divergent. They yield contributions
to the mass and field strength renormalization.
All the other diagrams are superficially convergent, they
do not depend on $\tilde{w}$ and 
yield no contribution to $\tilde{\Delta}_3$.
Here we only demonstrate the calculation of

\vspace{2cm}

{\bf Diagram 6:}

\begin{center}
\begin{picture}(70,30)(13,13)
\Vertex(20,50){2}
\Vertex(60,50){2}
\Vertex(100,50){2}
\Line(0,50)(20,50)
\Line(100,50)(120,50)
\CArc(40,50)(20,0,360)
\CArc(80,50)(20,0,360)
\CArc(60,50)(40,0,180)
\end{picture}
\end{center}

With $i=2, j_1=2$ and $j_2=1$ we obtain from (\ref{77})
$K(2,2,1) = \frac{(4!)^3}{4}$ and inserting this in (\ref{78})
we get the following expression:

\begin{eqnarray}
S^{(6)}_3 (g) & = & \frac{i \lambda ^3}{4!} \int d^4 x_1
\int d^4 x_2 \int d^4 x_3 \ : \phi (x_2) \phi (x_3): g(x_1)
g(x_2) g(x_3) \cdot \nonumber \\
& & \cdot \biggl[ \left( f_{12}^{(3)}+ f_3 ^{(3)}\right) \left( i \Delta_F
(x_1 - x_2) \right)^2 _R  \left( i \Delta_F (x_1 - x_3) \right)^2
i \Delta_F (x_2 - x_3) \nonumber \\
& & \quad
+  \left( f_{13}^{(3)}+ f_2 ^{(3)}\right) \left( i \Delta_F
(x_1 - x_2) \right)^2  \left( i \Delta_F (x_1 - x_3) \right)^2_R
i \Delta_F (x_2 - x_3) \nonumber \\
& & \quad
+  \left( f_{23}^{(3)}+ f_1 ^{(3)}\right) \left( i \Delta_F
(x_1 - x_2) \right)^2  \left( i \Delta_F (x_1 - x_3) \right)^2
(i \Delta_F (x_2 - x_3))_R  \biggr]_R \nonumber
\end{eqnarray}

At first we have to perform the superficial renormalization,
therefore we introduce new variables 
$u=x_2 - x_1, v=x_1-x_3$ and apply the operator $W^{(2)}(2, \tilde{w},
v, u+v)$ to the Wick-monomial and the test functions. Because of
remark 3 (before the calculations) the derivatives in the 
subtraction act only on the Wick-monomial depending only on one 
variable $u+v$. In this case $W^{(2)}$ has the form of a
subtraction operator in one variable given by (\ref{fortform}).

\begin{eqnarray}
S^{(6)}_3 (g) & = & \frac{i \lambda ^3}{4!} \int d^4 u
\int d^4 v \int d^4 x_3 \nonumber \\
& & \cdot \biggl[ \left( f_{12}^{(3)}+ f_3 ^{(3)}\right) \left( i \Delta_F
(u) \right)^2 _R  \left( i \Delta_F (v) \right)^2
i \Delta_F (u+v) \label{l1} \\
& & \quad +  \left( f_{13}^{(3)}+ f_2 ^{(3)}\right) \left( i \Delta_F
(u) \right)^2  \left( i \Delta_F (v) \right)^2_R
i \Delta_F (u+v) \label{l2} \\
& & \quad  +  \left( f_{23}^{(3)}+ f_1 ^{(3)}\right) \left( i \Delta_F
(u) \right)^2  \left( i \Delta_F (v) \right)^2
(i \Delta_F (u+v))_R  \biggr] \label{l3} \\
& & \cdot \biggl[ :\phi (u+v+x_3) \phi (x_3): g(u+v+x_3) g(v+x_3)
g(x_3) \nonumber \\
& & \quad - \tilde{w} (u) \tilde{w} (v) \tilde{w} (u+v) :\phi ^2 (x_3):
g^3 (x_3) \nonumber \\
& &  \quad - \tilde{w} (u) \tilde{w} (v) \tilde{w} (u+v) \frac{(u+v)^2}{8}
:\phi (x_3) \Box \phi (x_3): g^3 (x_3) \biggr]. \nonumber
\end{eqnarray}

Because of $(i \Delta_F (u+v))_R = i \Delta_F (u+v)$ we have now to
renormalize the subdivergences of the lines (\ref{l1}) and (\ref{l2}).
Using $ \left( f_{12}^{(3)}+ f_3 ^{(3)}\right)_{u=0} =1$ we obtain
from line (\ref{l1}):

\begin{eqnarray}
\lefteqn{\left( i \Delta_F (u) \right)^2 W^{(1)} (0,w,u)  
\left( f_{12}^{(3)}+ f_3 ^{(3)}\right)\left( i \Delta_F (v) \right)^2
i \Delta_F (u+v) \cdot} \nonumber \\
& &  \cdot \biggl[ :\phi (u+v+x_3) \phi (x_3): g(u+v+x_3) g(v+x_3)
g(x_3)
- \tilde{w} (u) \tilde{w} (v) \tilde{w} (u+v) :\phi ^2 (x_3):
g^3 (x_3) \nonumber \\
& & \quad - \tilde{w} (u) \tilde{w} (v) \tilde{w} (u+v) \frac{(u+v)^2}{8}
:\phi (x_3) \Box \phi (x_3): g^3 (x_3) \biggr] \nonumber \\
& = & \left( i \Delta_F (u) \right)^2 \left( i \Delta_F (v) \right)^2
\nonumber \\
& & \biggl\{ \left( f_{12}^{(3)}+ f_3 ^{(3)}\right) i \Delta_F (u+v)
\biggl[ :\phi (u+v+x_3) \phi (x_3): g(u+v+x_3) g(v+x_3)
g(x_3) \nonumber \\
& & \qquad \qquad  \qquad \qquad \qquad \qquad
- \tilde{w} (u) \tilde{w} (v) \tilde{w} (u+v) :\phi ^2 (x_3):
g^3 (x_3) \nonumber \\
& &  \qquad \qquad  \qquad \qquad \qquad \qquad
- \tilde{w} (u) \tilde{w} (v) \tilde{w} (u+v) \frac{(u+v)^2}{8}
:\phi (x_3) \Box \phi (x_3): g^3 (x_3) \biggr] \nonumber \\
& & \quad - i \Delta_F (v) \biggl[ :\phi (v+x_3) \phi (x_3): g^2 (v+x_3)
g(x_3) - \tilde{w}^2 (v) :\phi^2 (x_3) : g^3 (x_3) \nonumber \\
& & \quad \qquad \qquad
- \tilde{w}^2 (v) \frac{v^2}{8} : \phi(x_3) \Box \phi(x_3):
g^3(x_3) \biggr]  \biggr\} . \nonumber 
\end{eqnarray}

Interchanging $u$ and $v$ we obtain from this the renormalization
of the subdivergence of line (\ref{l2}). Finally we obtain the 
following result:

\begin{eqnarray}
S^{(6)}_3 (g)
& =  &  \frac{i \lambda ^3}{8}
\int d^4 u \int d^4 v \int d^4 x_3 \ \left( i \Delta_F (u) \right)^{2}
\left( i \Delta_F (v) \right)^{2}
\nonumber \\
& & \biggl\{
i \Delta_F (u+v)
\Bigl[ : \phi (x_3) \phi (u+v+x_3): g(u+v+x_3) g(v+x_3) g(x_3)
\nonumber \\ 
& & \qquad \qquad \qquad  
- \tilde{w}(u) \tilde{w}(v) \tilde{w}(u+v)  
: \phi ^2(x_3): g^3(x_3) \nonumber \\
& & \qquad \qquad \qquad  
- \tilde{w}(u) \tilde{w}(v) \tilde{w}(u+v) \frac{1}{8} \left( u+v \right)^2 
:\phi (x_3) \Box \phi (x_3): g^3(x_3) 
\Bigr]  \nonumber \\
& & - i \Delta_F (v) w(u) \Bigl[ 
: \phi (x_3) \phi (v+x_3): g^2(v+x_3) g(x_3)
\nonumber \\
& & \qquad \qquad \qquad 
- \tilde{w}^2(v)  : \phi ^2(x_3): g^3(x_3) \nonumber \\
& & \qquad \qquad \qquad 
- \tilde{w}^2(v) \frac{v^2}{8}
: \phi (x_3) \Box \phi (x_3): g^3(x_3) 
\Bigr]  \nonumber \\
& & - i \Delta_F (u)  w(v)
\Bigl[ 
: \phi (u+x_3) \phi (x_3): g(u+x_3) g^2(x_3)
\nonumber \\
& &  \qquad \qquad \qquad 
- \tilde{w}^2(u)  : \phi ^2(x_3): g^3(x_3) \nonumber \\
& &  \qquad \qquad \qquad 
- \tilde{w}^2(u) \frac{u^2}{8}
: \phi (x_3) \Box \phi (x_3): g^3(x_3) 
\Bigr] \biggr\}. \nonumber 
\end{eqnarray}

The renormalized diagrams 1-5 and 7-11 are given in the appendix
\ref{ergebnisse}.

Now it becomes clear that the independence of the partition of
unity proved in \cite{fredneu} is a consequence of the fact that the Taylor
subtractions of the subdivergences act on the partition of unity. 
In the explicit calculation we see that not only the $T$-products
as a whole as shown in \cite{fredneu} but also the contributions
to the individual diagrams are independent  of the partition of unity.

\subsection{The Normalization Conditions}

We have seen how renormalization works in the Epstein Glaser
formalism in $\Phi ^4$-theory. To come back to a more
theoretical formulation we present the normalization conditions
for $\Phi ^4$-theory. Some of them
are an abstract formulation of techniques
used in the previous calculations. 
In contrast to the calculations of the previous section the
following conditions are independent of the adiabatic limit.
The normalization conditions, introduced in \cite{mdf} and extended
in \cite{fm}, restrict the ambiguities in the renormalization.
We repeat these conditions in the more simple form
fitting to $\Phi^4$-theory:

\begin{itemize}

\item Condition N1 demands the Lorentz covariance of the $T$-products,
it is described in \cite{wir} and \cite{harakiri} how to realize
it in the extension procedure.

\item Condition N2 gives the form of the adjoint of $T$ on $\mathcal{D}$
and makes sure that the $S$-matrix is unitary:

\begin{eqnarray}
T_n(f_1, \ldots f_n)^+ = 
\sum_{P \in Part J} (-1)^{|P|+n} \prod_{p \in P} T_n \left( f_i^+,
i \in p \right), \nonumber
\end{eqnarray}

where the sum is over the ordered partitions of $J$.

\item It is shown in \cite{fm} that 
condition N3 is equivalent to the Wick expansion of the
timeordered products. 
The Wick expansion has the form

\begin{eqnarray}
T_n \left( g_1 \phi ^{l_1}, \ldots, g_n \phi ^{l_n} \right)
& = & \int \ dx_1 \ldots \int \ dx_n \sum_{0 \leq k_1, \ldots, k_n
\atop k_i < l_i}  <0| 
T \left( \phi^{k_1}(x_1) \ldots \phi^{k_n}(x_n) \right) |0> \nonumber \\
& & \prod_{i=1}^n \frac{l_i!}{k_i! (l_i-k_i)!}
: \phi^{l_1 - k_1}(x_1) \ldots \phi^{l_n - k_n}(x_n): g(x_1)
\ldots g(x_n). \label{wick}  
\end{eqnarray}

Each contribution to the sum in (\ref{wick}) corresponds to a diagram
with $\frac{1}{2} \sum_i k_i$ internal lines and $\sum_i (l_i -k_i)$
external lines. Condition N3 reads

\begin{eqnarray}
\lefteqn{\left[ T_n \left( f_1, \ldots, f_n \right),
T_1 \left( g \phi \right) \right] =} \nonumber \\ 
& & = \sum_{k=1}^n i T_n 
\left( f_1, \ldots , \Delta \frac{ \partial f_k}{\partial \phi}, \ldots
, f_n \right)
+  \sum_{k=1}^n i T_n 
\left( f_1, \ldots , 
\left( \partial _{\mu}\Delta \right) 
\frac{ \partial f_k}{\partial (\partial _{\mu}
\phi)}, \ldots
,f_n \right),
\end{eqnarray}

for all $f_i$ containing only fields and their first derivatives
with  

\begin{eqnarray}
\Delta (x_k) = \int d^4 y \ \Delta _{11} (x_k-y) g(y)
\end{eqnarray}

where $\Delta_{11}(y-x_k)$ is the commutatorfunction
$i\Delta_{11}(y-x_k) = [:\phi (y):, :\phi (x_k):]$. 

The derivative of $f$ is implicitly defined by N3 for $n=1$ for all 
$f \in \mathcal{D} \left( \R^4, \mathcal{A} \right)$ containing only
linear combinations of fields and their first derivatives.
We obtain

\begin{eqnarray}
\frac{ \partial}{\partial \phi}  \left( g_n (y) \phi^n \right) = n g_n (y)
\phi ^{n-1}, \qquad
\qquad
\frac{ \partial}{\partial \phi} 
\left( g(y) \partial _{\mu}\phi\right) = 0.
\label{ableitung1}
\end{eqnarray}

By demanding

\begin{eqnarray}
T_n \left( f_1, \ldots, f_n, g(z) (\Box + m^2) \phi \right)
= T_n \left( f_1, \ldots, f_n, ((\Box_z + m^2) g(z)) \phi \right)
\end{eqnarray}

we can extend condition N3 to $T$-products
containing one factor $(\Box + m^2) \phi$.

\item Condition N4 has the form

\begin{eqnarray}
T_{n+1} 
\left( f_1, \ldots, f_n,- \left( \Box_y + m^2 \right) 
g(y) \phi  \right)
= i \sum_{k=1}^n T_n
\left( f_1, \ldots , g \frac{ \partial f_k}{\partial \phi}, \ldots
, f_n \right) 
\end{eqnarray}

where the $f_i \in \mathcal{D} \left( \R^4, \mathcal{A} \right)$ 
contain only combinations of fields and their first derivatives.

We now show that in $\Phi^4$-theory the Dyson Schwinger equations
are a consequence of N4. With the Gell Mann Low formula the Green's 
functions $G_i \left( x_1, \ldots x_n \right)$ have the form

\begin{eqnarray}
\frac{
\left\langle 0 \left| T \phi(x_1) \ldots \phi (x_i) \sum_{n=0}^{\infty}
\frac{i^n}{n!} \int dy_1 \ldots \int dy_n \left( \frac{-\lambda}{4!}
\right)^n \phi ^4(y_1) \ldots \phi^4 (y_n) g(y_1) \ldots g(y_n) \right|
0 \right\rangle}{\left\langle 0 \left| T \sum_{n=0}^{\infty}
\frac{i^n}{n!} \int dy_1 \ldots \int dy_n \left( \frac{-\lambda}{4!}
\right)^n \phi ^4(y_1) \ldots \phi^4 (y_n) g(y_1) \ldots g(y_n) \right|
0 \right\rangle}.
\end{eqnarray}

So we obtain with N4 and (\ref{ableitung1})

\begin{eqnarray}
\lefteqn{ \int d^4 z \ 
\left( \Box_z + m^2 \right) G_i \left( x_1, \ldots x_{i-1},z \right)
=} \nonumber \\
& & = \int d^4 z \ \left[ i 
\sum_{k=1}^{i-1} \delta (x_k - z) G_{i-2} \left(x_1, \ldots 
\check{x}_k \ldots x_{i-1} \right) + 
\frac{\lambda g(z)}{6} 
G_{i+2} \left( x_1, \ldots x_{i-1},z,z,z \right) \right] \nonumber \\
\end{eqnarray}

and these are the Dyson Schwinger equations \cite{rivers}.
With $G_{i+2} \left( x_1, \ldots x_{i-1},z,z,z \right)$ we mean 
the vacuum expectation value of the fieldproduct
$\phi (x_1) \cdot \ldots \phi (x_{i-1}) \phi ^3 (z)$. 

\end{itemize}

\section{Topics Around the Action Principle}

\subsection{Main Theorem of Perturbative Renormalization Theory}
\label{mt}

The procedure of renormalization depends on the renormalization
scale, and we can say that two $T$-products $T$ and $\hat{T}$
are renormalizations at different scales $m$, although their 
scale dependence is not obvious in the abstract formulation of 
section \ref{tprod}.
Given two renormalization prescriptions  $T$ and $\hat{T}$
belonging to different scales $m$ one can pass from one to another
by a suitable change
of the scale dependent quantities in the Lagrangian.
This is the content of the main theorem of perturbative 
renormalization theory, which is proved in \cite{stora}
in the framework of causal perturbation theory.
Popineau and Stora give a construction of the changes of
parameters which is, due to the inductive form, not suited
for calculations in higher orders. In this section we derive the explicit
form of the changes of parameters
in the Lagrangian compensating a change of renormalization 
$T \to \hat{T}$, which corresponds to the structure of renormalization
given in \cite{zav}.
In the following calculations we use the abbreviations

\begin{eqnarray}
f & := & g(x) \phi ^4 \nonumber \\
T E_k (f) & := &  \sum_{n=k}^{\infty}
\frac{(-i)^n}{n!} 
T_n \left( f^{\bigotimes n} \right). \nonumber 
\end{eqnarray}

Then the $S$-matrix reads, for example, 

\begin{eqnarray}
S = T E_0 (f) =  \sum_{n=0}^{\infty}
\frac{(-i)^n}{n!} 
T_n \left( f^{\bigotimes n} \right), \qquad \mbox{and} \qquad
\Delta E_k (f) =  \sum_{n=k}^{\infty}
\frac{(-i)^n}{n!} 
\Delta_n \left( f^{\bigotimes n} \right).
\end{eqnarray}

The following theorem describes how to absorb a change in the
renormalization prescription $\hat{T} \to T$ in a change $f \to
f_r$ of the physical parameters in the Lagrangian:

\begin{Thm} \label{renth}
An $S$-matrix renormalized according to $\hat{T}$ can be expressed
by an $S$-matrix renormalized according to $T$ in the
following way:

\begin{eqnarray}
\hat{T} E_0 (f) = T E_0 (\Delta E_1(f)) =: T E_0 (f_r). \label{101}
\end{eqnarray}

\end{Thm}

{\bf Remark:} Because of $\hat{T} E_0 (f) = 1 + T(f) + \hat{T} E_2 (f)$
and $\Delta E_1 (f) = f + \Delta E_2 (f)$
we can derive from (\ref{101}) the following
recursion relation:

\begin{eqnarray}
T \Delta E_2 (f) = \hat{T} E_2 (f) - T E_2 ( f+\Delta E_2 (f)).
\label{rec1}
\end{eqnarray}

{\bf Proof:} We prove the theorem in the following calculation:

\begin{eqnarray}
\hat{T} E_0 (f) = \sum_{n=0}^{\infty} \frac{(-i)^n }{n!} 
\hat{T} \left( f^{\otimes n} \right) 
\stackrel{(\ref{msatz})}{=}  \sum_{n=0}^{\infty} \frac{(-i)^n}{n!}
\sum_{P \in Part(J)}
T \left( \bigotimes_{O_i \in P}
\Delta \left( f^{\otimes |O_i|} \right) \right)
 . \nonumber \\
\end{eqnarray}

Here we sum over all partitions of the set $J = \{1
, \ldots , n\}$. Now we take a fixed partition $P$
and denote with $N_i$ the number of elements $O$ of $P$
with $|O|=i$.

Then the relation $n= \sum_{i=1}^k i N_i$
is true with $k \leq n$. There are

\begin{eqnarray}
\frac{ n!}{N_1 ! \ldots N_k ! 1!^{N_1} \ldots k! ^{ N_k}}
\end{eqnarray}

different partitions $P$ with the same numbers $N_i$.
Therefore we have

\begin{eqnarray}
\hat{T} E_0 (\V) & = & \sum_{n=0}^{\infty} \frac{(-i)^n }{n!}
\sum_{\sum i N_i = n}
\frac{n!}{N_1 ! \ldots N_k ! 1!^{N_1} \ldots k! ^{ N_k}}
T_n \Biggl[ \left( \Delta  (f) \right)^{N_1}
\ldots 
\left( \Delta_k  (f^{\bigotimes k}) \right)^{N_k} \Biggr] \nonumber \\
\end{eqnarray}

and obtain

\begin{eqnarray}
\hat{T} E_0 (f) & = & \lim _{k \to \infty} T \Biggl[
\sum _{N_1 =0}^{\infty}  \frac{(-i)^{N_1} }{N_1!}
\left( \Delta  (f)  \right)^{N_1} \cdot 
\sum _{N_2 =0}^{\infty}  \frac{(-i)^{2N_2} }{N_2!2!^{N_2}}
\left( \Delta  (f \otimes f) \right)^{N_2} \cdot \ldots
\nonumber \\
& & \qquad \qquad \ldots \cdot 
\sum _{N_k =0}^{\infty} \frac{(-i)^{kN_k} }{N_k!k!^{N_k}}
\left( \Delta_k   (f^{\bigotimes k} \right)^{N_k} \Biggr] 
 = T \left( E_0 \left( \Delta E_1 (f) \right) \right) 
\qquad  \blacksquare
\end{eqnarray}

In renormalizable theories, $f_r = \Delta E_1 (f)$ consists only
of linear combinations of the field combinations in the original 
Lagrangian. In scalar $\Phi ^4$ -theory $\Delta E_1 (f)$ is
a linear combination of the monomials
$\partial _{\mu} \Phi  \partial ^{\mu} \Phi,
\Phi ^2$ and $\Phi ^4$. 
The coefficients depend on the renormalization conditions
and this dependence is described by the Callan-Symanzik equations
and the renormalization group equations. The most elegant way
to derive them uses the action principle \cite{low}.
In the formalism of Epstein and Glaser the partition of the 
Lagrangian into a free and an interacting part is important.
The free part of the Lagrangian defines the Hilbert space, fields
and their masses of the free theory. It is not clear that a
contribution of a massterm to the interacting Lagrangian
$\Delta E_1 (f)$ can be interpreted as a shift in the mass,
the result of the action principle is that concerning the $S$-matrix
it can be interpreted as a massterm.

\subsection{Insertions}

To derive the action principle we need the notion of an insertion.

\begin{Def}
An insertion of degree $a$ of an element $g \in \mathcal{D} \left(
\R^4, \mathcal{A} \right)$ into a $T$-product is defined by

\begin{eqnarray}
I^{(a)}\left(  g ,T_n  \left( \mbox{$ \bigotimes _{j \in J}$} f_j \right) 
\right)
& := & T_{n+1} \left( 
\mbox{$ \bigotimes_{j \in J}$} f_j \otimes g^{(a)} \right). 
\end{eqnarray}

where $g^{(a)}$ means that the vertex belonging to $g$ is treated
in the extension procedure defining $T_{n+1}$ as a vertex 
of mass dimension $a$.

\end{Def}

{\bf Remark:} If the degree $a$ of the insertion corresponds to its
physical mass dimension $M$ it is called a soft insertion. In the case
$a>M$ all diagrams containing the vertex of the insertion are
oversubtracted and we have a hard insertion.
The relation of two insertions of the same Wick monomial
with different degrees is given by the Zimmermann Identities
derived in the next section. 

The behaviour of an insertion into the $S$-matrix 
under a change of renormalization
is given by the following theorem.

\begin{Thm} \label{insthm}

An insertion in the $S$-matrix renormalized according to $\hat{T}$
can be expressed as an insertion in the $S$-matrix renormalized
according to $T$ in the following way:

\begin{eqnarray}
I^{(a)} \left( g,
\hat{T} E_0 (f) \right) & = & I^{(a)} \left( \Delta (g E_0 (f)),
T E_0 (f_r) \right) \nonumber \\
& =: & I^{(a)} \left( g_r (f), 
T E_0 (f_r) \right). \label{ins}
\end{eqnarray}
\end{Thm}

{\bf Remark:} Analogously to (\ref{rec1}) we obtain
after some calculation the recursion relation
$T \Delta (g E_1 (f)) = I^{(a)} \left( g, \hat{T}E_1 (f) \right)
-  I^{(a)} \left( g, T E_1 (f_r) \right) 
-  I^{(a)} \left( \Delta (g E_1 (f) ),
T E_1 (f_r) \right).$

\vspace{1cm}

{\bf Proof:} We know from Theorem (\ref{renth})

\begin{eqnarray}
\hat{T} E_0 (h) = T E_0 (\Delta E_1(h)). \label{109}
\end{eqnarray}

Let $a$ be the degree of the insertion $g$, $\lambda$ a parameter
and $h= f + g \lambda$ such that the vertex $g \lambda$
is treated in the subtraction procedure of the $T$-products as
a vertex of massdimension $a$. Then we can interchange in the
following calculation differentiation and integration:

\begin{eqnarray}
I^{(a)} \left( g, \hat{T} E_0( f) \right) 
& = & i \frac{\partial}{\partial \lambda} \left. \hat{T} E_0 (h)
\right|_{\lambda =0} \nonumber \\
& \stackrel{(\ref{109})}{=} & i \frac{\partial}{\partial \lambda}
T \left( \left.
E_0 \left( \Delta E_1 (f + 
g \lambda) \right) \right|_{\lambda =0} \right) \nonumber \\
& = & I^{(a)} \left(  g \Delta (E_0 (f)),
T E_0 (\Delta E_1(f)) \right).
\blacksquare
\end{eqnarray}

The following properties of insertions are interesting:

\begin{itemize}

\item Interacting fields have up to a factor $S^{-1}$ 
the form of an insertion:

\begin{eqnarray}
\int d^4 x A_{int \mathcal{L}}(x) & = & \frac{\delta}{\delta h(x)}
S_T (\mathcal{L})^{-1} S_T (\mathcal{L} +hA) |_{h=0}
\nonumber \\
& = & S_T (\mathcal{L})^{-1} I\left( A, S_T (\mathcal{L})\right).
\end{eqnarray}

\item Insertions in the $S$-matrix
of monomials $f_j$ occuring in the
interaction act as counting operators of vertices of the
kind  $f_j$. The proof is analogously to the one given in
\cite{low}:
The form of the $S$-matrix of a theory with interaction vertices
$f_1, \ldots f_n$ of mass dimensions $a_1, \ldots a_n$,

\begin{eqnarray}
S_T & = & T E_0 \left( f_1^{(a_1)}+ \ldots +  f_n^{(a_n)} \right)
\nonumber \\
& = & \sum_{c_i = 0}^{\infty} \frac{(-i)^{\sum c_i}}
{c_1! \ldots c_n!}
T_{c_1 + \ldots +c_n} \left( \left( f_1^{(a_1)} \right) ^{\otimes c_1}
\ldots  \left( f_n^{(a_n)} \right) ^{\otimes c_n} \right), 
\end{eqnarray}

implies that the contribution of a diagram $\gamma$ with $c_i$
vertices of the kind $f_i$ to the $S$-matrix has the form

\begin{eqnarray}
S_{(c_1, \ldots,c_n)} =  \frac{(-i)^{\sum c_i}}{c_1! \ldots c_n!} 
T_{c_1 + \ldots +c_n} 
\left( \left( f_1^{(a_1)} \right) ^{\otimes c_1}
\ldots  \left( f_n^{(a_n)} \right) ^{\otimes c_n} \right).
\end{eqnarray}

An insertion of a vertex $f_j$ of degree $a_j$ in the $S$-matrix
yields

\begin{eqnarray}
I^{(a_j)} \left( f_j, S_T \right) & \stackrel{def.}{=} &
I^{(a_j)} \left( f_j,
T E_0 \left( f_1^{(a_1)}+ \ldots +  f_n^{(a_n)} \right)
\right) \nonumber \\
& = & T \left( \sum_{c_i = 0}^{\infty} \frac{c_j
(-i)^{\sum c_i}}{c_1! \ldots c_n!}
\left( f_1^{(a_1)} \right) ^{\otimes c_1} 
\ldots  \left( f_n^{(a_n)} \right) ^{\otimes c_n}
\right) \nonumber \\
& = & \sum_{c_i = 0}^{\infty} c_j S_{(c_1, \ldots,c_n)}.
\end{eqnarray}

\item Zimmermann Identities

The Zimmermann identities \cite{zim} are relations of insertions of different
degrees.
Comparing the two insertions

\begin{eqnarray}
I ^{(a)} \left( g ,T \left( \mbox{$ \bigotimes_{j \in J}$} 
f_j \right) \right) 
& = & T \left( \mbox{$ \bigotimes_{j \in J}$} 
f_j \otimes g \right), \nonumber \\
I^{(b)} \left( g,T \left( \mbox{$ \bigotimes_{j \in J}$} 
f_j \right) \right) 
& = & T^* \left( \mbox{$ \bigotimes_{j \in J}$} f_j \otimes g \right)
\end{eqnarray}

we obtain the operatorvalued distribution $\Delta$ mediating between the
two $T$-products.
Using the properties

\begin{eqnarray}
\Delta \left( f_i \right) & = & f_i, \nonumber \\
\Delta \left( g \right) & = & g, \nonumber \\
\Delta \left( \bigotimes_{i \in J} f_i \right) & = & 0
\ \mbox{for} \ |J|>1
\end{eqnarray} 

we obtain with (\ref{msatz}) the Zimmermann identities 

\begin{eqnarray}
T^* \left( g \otimes \bigotimes_{i \in J} f_i \right) = 
T \left( g \otimes  \bigotimes_{i \in J} f_i \right) 
+ \sum_{O_k \subseteq J \atop O_k \not= \emptyset} 
T \left( \bigotimes_{l\not\in O_k} f_l \otimes
\Delta \left(  \bigotimes_{l \in O_k} f_l \otimes g \right) \right).
\end{eqnarray}

In the special case of 
insertions in the $S$-matrix this formula simplifies to

\begin{eqnarray}
I^{(b)} \left( g, TE_0(f) \right)
& \stackrel{(\ref{ins})}{=} & I^{(a)}
\left( \Delta(g E_0(f)), T(E_0(f) \right)
\nonumber \\
& = & I^{(a)} \left( g, T E_0(f) \right) 
+ I^{(a)} \left( \Delta(g E_1(f)),
T E_0(f) \right). \label{trans}
\end{eqnarray}

\item The Action Principle

The action principle describes the effects of a change of
parameters of a theory \cite{sibold}. 
Lowenstein \cite{low} has formulated it in terms of insertions
into the $S$-matrix, and we follow his derivation. In the adiabatic limit
it can be transformed with the Gell-Mann-Low formula into the
usual formulation in Green's functions. 

According to the main theorem (\ref{mt}) a change of the 
renormalization $T$ can be absorbed by finite counterterms shifting 
the quantities in the Lagrangian. Therefore we can assume without
loss of generality that the $T$-product is fixed. 
A change of a parameter $p \to p + \delta p$ can cause changes
in the free Lagrangian $\mathcal{L}_0  
 \rightarrow  \mathcal{L}_0 + \delta \mathcal{L}_0$
and in the interaction part of the
Lagrangian:$f  \rightarrow  f+ \delta f$.
The action principle states that a change of the 
$S$-matrix caused by a variation of a parameter $p$ can be expressed
by an insertion in the $S$-matrix. We only regard the case
where only the interaction part $f$ depends on the parameter $p$.
In this case the $T$-product is independent of $p$ and we obtain
by a trivial differentiation the first part of the action principle:

\begin{eqnarray}
\frac{\partial}{\partial p}  T \left( E_0 (f) \right)
= I^{(4)} \left( \frac{\partial f}{\partial p}, T E_0 (f) 
\right). \label{wip}
\end{eqnarray}

\end{itemize}

\section{Summary and Outlook}

We have derived the structure of finite renormalizations in the
Epstein Glaser formalism. The form
of the $S$-matrix after a finite renormalization and the
Zimmermann identities follow from this result.
We showed that in $\phi^4$ theory the Dyson Schwinger
Equations are a consequence of N4.

Furthermore we have performed the renormalization of the
$S$-matrix in $\Phi ^4$-theory up to third
order. The result depends on the testfunction $w$ used in the
renormalization, and finite counterterms can be read off.
Comparing this procedure with other renormalizations in momentum
space, we can say that Epstein Glaser renormalization is better
suited for the treatment of diagrams with few vertices and many loops, 
whereas momentum space renormalization has advantages in the
renormalization of diagrams with many vertices and few loops.

Finally we defined insertions into $T$-products. 
The part of the action principle concerning changes in the interaction
is formulated in the Epstein Glaser formalism
independent of the adiabatic limit.

There are still many open questions: the other part of the
action principle has to be derived and a fully translation
of the derivation of the Callan Symanzik equations as in
\cite{low} has to be worked out.
An interpretation of changes of masses has to be found in
the Epstein Glaser framework.
One has to check further if there appear problems in the
translation of these methods on curved space times.

\vspace{1cm}

{\bf Acknowledgements:} I thank Prof. K. Fredenhagen for many
discussions and the DFG for financial support.
The referee is acknowledged for his valuable comments
on the manuscript and his patience.

\vspace{1cm}

{\LARGE \bf Appendix}

\begin{appendix}

\section{Results of the Third Order Calculations} \label{ergebnisse}

The contributions of the diagrams 1 and 7 vanish in the adiabatic limit.

Diagrams 2, 3 and 4 are superficial logarithmic divergent,
they have four external legs and contribute to the renormalization
of the coupling constant. The calculation of diagram 2 is given in the
appendix. Here we list only the results, weightened with $N$:

\vspace{2cm}

{\bf Diagram 2:}

\begin{center}

\begin{picture}(70,30)(13,13)
\Vertex(20,50){2}
\Vertex(60,50){2}
\Vertex(100,50){2}
\Line(0,50)(120,50)
\CArc(40,50)(20,0,360)
\end{picture}
\end{center}

\begin{eqnarray}
S^{(2)}_3 (g) & = &
\frac{i \lambda ^3}{36}
\int d^4 u \int d^4 v \int d^4 x_3
\left( i \Delta_F (u) \right)^3 i \Delta_F (v)
 \nonumber \\
& & \Bigl\{
 : \phi ^3 (x_3) \phi (u+v+x_3):  g(u+v+x_3)   g(v+x_3) g(x_3) 
\nonumber \\ 
& & \quad  
- \tilde{w}(u) \tilde{w}(v) \tilde{w}(u+v)
: \phi ^4 (x_3): g^3(x_3) \nonumber \\
& & \quad  
-w(u)  : \phi ^3 (x_3) \phi (v+x_3):  g^2(v+x_3) g(x_3) 
\nonumber \\
& & \quad 
+ w(u) \tilde{w}^2 (v) :\phi ^4 (x_3): g^3(x_3) \nonumber \\
& & \quad  
- \frac{w(u)}{8} u^2 : \phi ^3 (x_3) \Box \phi (v+x_3):
g^2(v+x_3) g(x_3) \nonumber \\
& & \quad  
+ \frac{w(u)}{8} u^2 \tilde{w}(v) \left( \Box \tilde{w}(v) \right)
: \phi ^4 (x_3): g^3(x_3) \Bigr\}.
\end{eqnarray}

\vspace{1cm}

{\bf Diagram 3:}

\begin{center}
\begin{picture}(70,30)(13,13)
\Vertex(20,50){2}
\Vertex(60,50){2}
\Vertex(100,50){2}
\Line(0,40)(20,50)
\Line(0,60)(20,50)
\Line(100,50)(120,60)
\Line(100,50)(120,40)
\CArc(40,50)(20,0,360)
\CArc(80,50)(20,0,360)
\end{picture}
\end{center}

\begin{eqnarray}
S^{(3)}_3 (g) & = &
\frac{i \lambda ^3}{32}
\int d^4 u \int d^4 v \int d^4 x_3
\left( i \Delta_F (u) \right)^2 \left(i \Delta_F (v)\right)^2
 \nonumber \\
& & \Bigl\{
 : \phi ^2 (v+u+x_3) \phi ^2 (x_3):  g(v+u+x_3)   g(v+x_3) g(x_3) 
\nonumber \\ 
& & \quad 
- \tilde{w}(u) \tilde{w}(v)  \tilde{w}(u+v)  
:\phi ^4 (x_3):  g^3(x_3) \nonumber \\
& & \quad 
-w(u)  : \phi ^2 (v+x_3) \phi ^2 (x_3):  g^2(v+x_3) g(x_3) 
\nonumber \\
& & \quad 
+w(u) \tilde{w}^2(v)  :\phi ^4 (x_3):  g^3(x_3) \nonumber \\
& & \quad
-w(v)  : \phi ^2 (x_3+u) \phi ^2 (x_3):  g(x_3+u) g^2(x_3) 
\nonumber \\
& & \quad 
+w(v) \tilde{w}^2(u)  :\phi ^4 (x_3):  g^3(x_3)  \Bigr\}.
\end{eqnarray}

\vspace{2cm}

{\bf Diagram 4:}

\begin{center}
\begin{picture}(70,30)(13,13)
\Vertex(20,60){2}
\Vertex(80,60){2}
\Vertex(50,30){2}
\Line(0,60)(100,60)
\CArc(50,60)(30,0,360)
\Line(50,30)(40,20)
\Line(50,30)(60,20)
\end{picture}
\end{center}

\begin{eqnarray}
S^{(4)}_3 (g) & = & \frac{i \lambda ^3}{6}
\int d^4 u \int d^4 v \int d^4 x_3 \nonumber \\
& & \biggl\{ \left( i \Delta_F (u) \right)^{2}
 i \Delta_F (u+v)  i \Delta_F (v) \cdot \nonumber \\
& & \qquad
\cdot \Bigl[ : \phi ^2 (x_3) \phi (v+x_3) \phi (u+v+x_3):
g(u+v+x_3)g(v+x_3) g(x_3)  \nonumber \\
& & \qquad \qquad
- \tilde{w}(u) \tilde{w}(v) \tilde{w}(u+v) 
: \phi ^4 (x_3): g^3 (x_3) \Bigr]  \nonumber \\
& & - \left( i \Delta_F (u) \right)^{2}
\left( i \Delta_F (v) \right)^{2}  w(u)
\Bigl[ : \phi ^2 (x_3) \phi ^2 (v+x_3):
g^2(v+x_3) g(x_3)  \nonumber \\
& & \qquad \qquad  \qquad \qquad \qquad \qquad
- \tilde{w}^2 (v) : \phi ^4 (x_3): g^3 (x_3) \Bigr] \biggr\}.  
\end{eqnarray}

We see that $\tilde{w}$ indeed appears only in combination with the 
Wick monomial $:\phi ^4 (x_3):$.

Diagram 5 and 6 have two external lines and are therefore
superficial quartic divergent. They yield contributions
to the mass and field strength renormalization:

\vspace{2cm}

{\bf Diagram 5:}

\begin{center}
\begin{picture}(70,30)(13,13)
\Vertex(30,50){2}
\CArc(60,50)(30,0,360)
\CArc(90,50)(16,0,360)
\Vertex(85,35){2}
\Vertex(85,65){2}
\Line(30,50)(20,40)
\Line(30,50)(20,60)
\end{picture}
\end{center}

\begin{eqnarray}
S^{(5)}_3 (g)
& = &  \frac{i \lambda ^3}{24}
\int d^4 u \int d^4 v \int d^4 x_3  \left( i \Delta_F (u) \right)^{3}
 i \Delta_F (v)
\nonumber \\
& & \biggr\{ i \Delta_F (u+v)   
\Bigl[ :\phi ^2 (x_3): g(u+v+x_3)g(v+x_3) g(x_3) 
\nonumber \\
& & \qquad \qquad \qquad
- \tilde{w}(u) \tilde{w}(v) \tilde{w}(u+v) :\phi ^2 (x_3): g^3 (x_3) \Bigr] 
\nonumber \\
& & - i \Delta_F (v) 
w(u) \Bigl[ :\phi ^2 (x_3): g^2(v+x_3) g(x_3) \nonumber \\
& & \qquad \qquad \qquad
- \tilde{w}^2(v) :\phi ^2 (x_3): g^3 (x_3) \nonumber \\
& & \qquad \qquad \qquad
- \frac{u^2}{8} \left( \Box \tilde{w}(v)
\right) \tilde{w}(v) :\phi ^2 (x_3): g^3 (x_3)
\Bigr] \nonumber \\
& & - \frac{w(u)}{8} u^2  \left( \Box  i \Delta_F (v) \right)
\Bigl[  :\phi ^2 (x_3): g^2(v+x_3) g(x_3) \nonumber \\
& & \qquad \qquad \qquad \qquad \qquad
- \tilde{w}^2(v)  :\phi ^2 (x_3): g^3 (x_3)\Bigr]  \nonumber \\
& & + \frac{w(u)}{8} u^2
\left( \partial _{\mu}  i \Delta_F (v) \right)
\left( \partial ^{\mu} \tilde{w}(v) \right)
\tilde{w}(v) :\phi ^2 (x_3): g^3 (x_3) 
\biggr\}.
\end{eqnarray}

\vspace{2cm}

All the other diagrams are superficial convergent, but some of them 
contain subdivergences.
We obtain

{\bf Diagram 8:}

\begin{eqnarray}
S^{(8)}_3 (g)
& =  &  \frac{i \lambda ^3}{288}
\int d^4 u \int d^4 v \int d^4 x_3 \ \left( i \Delta_F (u) \right)^{3}
\nonumber \\
& & \Bigl\{ : \phi ^4 (x_3) \phi (v) \phi (u+v): g(u+v) g(v) g(x_3)
\nonumber \\
& & - w(u) 
: \phi ^4 (x_3) \phi ^2(v) :  g^2(v) g(x_3) \nonumber \\
& & 
- \frac{w(u)}{8} u^2 : \phi ^4 (x_3) \left( \Box
\phi (v) \right)  \phi (v): g^2(v) g(x_3) \Bigr\}, \nonumber \\
\end{eqnarray}

{\bf Diagram 9:}

\begin{eqnarray}
S^{(9)}_3 (g)
& =  &  \frac{i \lambda ^3}{24}
\int d^4 u \int d^4 v \int d^4 x_3 \ \left( i \Delta_F (u) \right)^{2}
 i \Delta_F (v)
\nonumber \\
& & \Bigl\{  : \phi ^3 (x_3) \phi^2 (u+v+ x_3) 
\phi (v +x_3): g(u+v+x_3) g(v+x_3) g(x_3)
\nonumber \\
& & -  w(u) : \phi ^3 (x_3) \phi^3 (v):
 g^2(v) g(x_3) \Bigr\}, \nonumber \\
\end{eqnarray}

{\bf Diagram 11:}

\begin{eqnarray}
S^{(11)}_3 (g)
& =  &  \frac{i \lambda ^3}{384}
\int d^4 u \int d^4 x_2 \int d^4 x_3 \ \left( i \Delta_F (u) \right)^{2}
\nonumber \\
& & \Bigl\{  : \phi ^4 (x_3) \phi^2 (x_2) \phi ^2 (u+x_2): 
g(u+x_2) g(x_2) g(x_3)
\nonumber \\
& & -  w(u) : \phi ^4 (x_3) \phi^4 (x_2):
 g^2(x_2) g(x_3) \Bigr\}. \nonumber \\
\end{eqnarray}

The last three diagrams are not dependent on $\tilde{w}$ and 
yield no contribution to $\tilde{\Delta}_3$.

\end{appendix}

\end{document}